\documentclass[manuscript]{aastex}
\shorttitle{EUV Spectra of Solar Flares from SPIRIT}
\shortauthors{S.~Shestov, S.~Kuzin et al.}

\newcommand{\cmt}{cm$^{-3}$}

\begin{document}

\title{EUV Spectra of Solar Flares from the EUV Spectroheliograph SPIRIT aboard
CORONAS-F satellite}

\author{S. Shestov and A. Reva\altaffilmark{1} and S. Kuzin}
\affil{Lebedev Physical Institute, Russian Academy of Sciences}
\email{sshestov@gmail.com}
\altaffiltext{1}{Moscow Institute of Physics and Technology (State University)}

\begin{abstract}
We present detailed EUV spectra of 4 large solar flares: M5.6, X1.3, X3.4, and
X17 classes in the spectral ranges 176--207~\AA\ and 280--330~\AA. These spectra
were obtained {by the slitless} spectroheliograph SPIRIT aboard the
CORONAS-F satellite.  To our knowledge these are the first detailed EUV spectra
of large flares obtained with spectral resolution of $\sim 0.1$~\AA. We
performed a comprehensive analysis of the obtained spectra and provide
identification of the observed spectral lines. The identification was performed
based {on the calculation} of synthetic spectra (CHIANTI database 
was used), with simultaneous {calculations of DEM} and density of the
emitting plasma. More than 50 intense lines are present in the spectra that
correspond to a temperature range of $T=0.5-16$~MK; most of the lines belong to
Fe, Ni, Ca, Mg, Si ions. In all the considered flares intense hot lines from
\ion{Ca}{17}, \ion{Ca}{18}, \ion{Fe}{20}, \ion{Fe}{22}, and \ion{Fe}{24} are
observed. The calculated DEMs have a peak at $T \sim 10$~MK. The densities were
determined using \ion{Fe}{11}--\ion{Fe}{13} lines and averaged $6.5 \times
10^9$~\cmt. We also discuss the identification, accuracy and major discrepancies
of the spectral line intensity prediction.
\end{abstract}


\section{Introduction}
The extreme ultra-violet (EUV) {emissions} of the solar corona have been studied
since the beginning of the space era due {to} the rich informational content of the
registered spectra. Analysis of such spectra allows the determination of various
plasma characteristics, such as temperature and density, and provides information
about {dynamic} processes that take place in the solar corona. In addition, the EUV
spectra of different coronal phenomena have become a subject of interest in {
a} number of different areas such as atomic physics, astrophysics and physics of
plasma.

Numerous spectroscopic observations have been carried out using spectroscopic
instruments of different types: slit spectroghraphs with high spatial
resolution, such as SERTS~\citep{Neupert92}, CDS/SOHO~\citep{Harrison95}, 
EIS/Hinode~\citep{Culhane2007}; spectroheliographs with imaging capabilities,
such as S082A/Skylab~\citep{Tousey77}, SPIRIT/CORONAS-F~\citep{Zhitnik2002}; and
full-Sun spectrographs, which obtain spectra from the whole solar disk, such as
those on {the} Aerobee rocket~\citep{Malinovsky73} or EVE/SDO~\citep{Woods2012}.

Data obtained in these experiments have been used for various goals {such
as} for {development} of atlases of spectral lines, validation of atomic data, 
measurement of temperature and density of the emitting plasma in different structures,
determination of presence of up- or downflows etc. Among the structures that
were studied, there are quiet sun regions~\citep{Brosius96}, active {regions}
(AR) cores~\citep{Tripathi2011}, off-limb AR plasma~\citep{ODwyer2011}, AR
mosses~\citep{Tripathi2010}, coronal streamers~\citep{Parenti2003}, bright
points~\citep{Ugarte-Urra2005} and others.

Whereas solar flares have also been observed by spectrographs, obtaining EUV
spectra of solar flares is not so common.  The first systematic analysis of EUV
flaring spectra was presented by \citet{Dere78}. The author analyzed  more than
50 photographic plates from the S082A spectroheliograph on Skylab and
constructed a catalog of spectral lines in the range 171--630~\AA. The catalog
included relative intensities of more than 200 spectral lines. 

Systematic studies of EUV spectra of solar flares have been continued on
subsequent satellites: SOHO (launched in 1995), Hinode (launched in 2006), 
{and} SDO (launched in 2010). The CDS spectrograph aboard {the} SOHO
satellite registered several large solar flares during their decay phases. The
first analysis of a CDS flare was made by \citet{Czaykowska99}. The
authors analysed intensities of spectral lines during the decay phase of a
M6.8 flare and determined density and temperature of post-flare loops.
\citet{DelZanna2006} also performed analysis of spectra {of a X17 flare
during the decay phase}. The authors studied Doppler shifts and found them to be
consistent with those, predicted {by a simple hydrodynamics model}.  It is
worth noting that due to the telemetry constraints of CDS, all these flares were
observed in fast-rastering regime in {only } 6 narrow spectral windows,
covering only a small portion of the wide spectral ranges 308--381 and
513--633~\AA\ of CDS. 

The EIS spectrograph aboard {the} Hinode satellite used an improved
optical layout with high efficiency EUV optics and detectors. Therefore, EIS has
superb spectral, spatial and temporal resolution {as well as} higher
telemetry volumes, which allow spectra { to be investigated} with much higher
details, such as {a} wider set of spectral lines, higher cadence,
{and} higher spatial resolution. EIS has observed a large span of flares,
starting from { small B2 class \citep{DelZanna2011} to large M1.8
\citep{Doschek2013}}. However, despite all its advantages, EIS usually observes
flares in coarse rastering regime. This fact limits the number of spectral lines
observed in a flarer; for example \citet{Watanabe2010} used only 17 lines for
plasma diagnostics from the whole spectral range 170--210 and 250--290~\AA.
There is a case when EIS has registered a full CCD flare spectrum
\citep{Doschek2013}; however, the authors focused on Doppler shift analysis and
used only 17 out of 500 lines registered by EIS.

The EVE spectrometer aboard SDO builds whole-Sun spectra in the range 10--1050
\AA. It has moderate spectral resolution of  1~\AA\, but operates with an
unprecedented 10 seconds cadence and almost 100~\% duty cycle. There are two
main difficulties in analysis of the EVE spectra: it has no spatial resolution
--- flare spectrum is mixed with the spectrum from the rest of the Sun, and due
to moderate spectral resolution of EVE most of the lines are blended. Despite
these obstacles EVE is widely used in solar investigations: for study of thermal
evolution of flaring plasma \citep{Chamberlin2012}, Doppler shifts study
\citep{Hudson2011}, and high temperature plasma electron density diagnostics
\citep{Milligan2012ne}.

Without diminishing the importance of the information obtained in these
experiments, it should be noted that a small number of EUV spectra of solar flares
have been registered so far, and published catalogs of spectral lines are
limited. 

In this paper we take advantage of the SPIRIT EUV spectroheliograph and perform
a comprehensive analysis of EUV spectra of four large solar flares. The flares
of M5.6, X1.3, X3.4 and X17 classes have been observed by a slitless EUV
spectroheliograph SPIRIT aboard CORONAS-F satellite. The spectroheliograph
operated in two wavelength ranges 176--207 and 280--330~\AA\ and {had a
spectral resolution} of 0.1~\AA.  We perform an absolute calibration of SPIRIT
spectral fluxes using simultaneous EIT/SOHO images. In order to identify the
obtained spectra we use an original approach, based on calculation of synthetic
spectra and its subsequent modification to match the observational data.
Simultaneously, we calculate DEM and $n_e$ of the emitting plasma and repeat
iteratively the whole procedure of identification several times.

We provide identification of {more than} 50 spectral lines in each
spectral band for each flare. In addition to spectral line intensities, we
{calculate the DEM} and plasma density for each flare. The obtained
information can be used not only for modelling of spectral fluxes in different
EUV spectral bands and for refinement of the atomic data, but also for studying
flares {themselves and validating} models for flare plasma evolution.

%% as many as 70 spectral lines of each spectral
%% band. Besides intensities of spectral lines, we calculate DEM and assess plasma
%% density in each flare. 
 
The obtained spectra, synthetic spectra, DEMs, and proposed IDL software are
available at \url{http://xras.lebedev.ru/SPIRIT/} or on request from S. Shestov.

\section{Observations}
\label{SPIRIT}
The SPIRIT complex of instrumentation was launched aboard the CORONAS-F
satellite \citep{Oraevskii2002e} on 31st July 2001 from Plesetsk cosmodrome,
northern Russia. The satellite was placed on a near-polar orbit with an
inclination of $82^{\circ}$ and a perigee of 500 km. The satellite carried 12
scientific instruments for the measurement of both particle and electromagnetic
emission of the Sun. The SPIRIT instrumentation was developed in
{the} Lebedev Physical Institute of the Russian Academy of Sciences and consisted of
telescopic and spectroheliographic channels for observation of the solar corona in
different soft X-ray and EUV spectral bands \citep{Zhitnik2002}. 

\begin{figure}[!h]
    \epsscale{.50} 
    \plotone{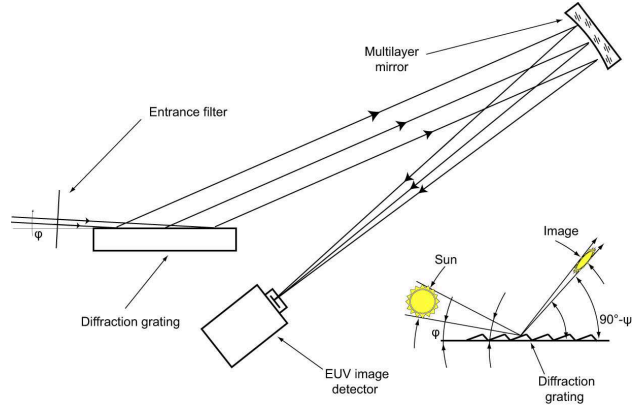}
    \caption{Optical scheme of EUV slitless spectroheliograph SPIRIT aboard CORONAS-F
    satellite.}
    \label{optical_scheme}
    \epsscale{1.0} 
  \end{figure}

The EUV spectroheliograph SPIRIT consisted of two similar independent spectral
channels: {\bf V190} channel for 176--207~\AA\ range, and {\bf U304} channel for
280--330~\AA\ range. Both channels were built using a slitless optical scheme
(see Figure~\ref{optical_scheme}). The solar EUV emission enters through an
entrance filter, falls on a  diffraction grating (with a grazing angle $ \varphi
\sim 1.5^{\circ}$). Diffracted radiation is focused on a detector by a mirror
with multilayer coating.  

The slitless optical scheme observes full-Sun FOV on the detector, 
which allowed us to obtain as many as 30 spectroheliograms with large solar flares during 4.5
years of the satellite's lifetime.

  %\placefigure{optical_scheme}

\begin{figure}[!h] 
    \plotone{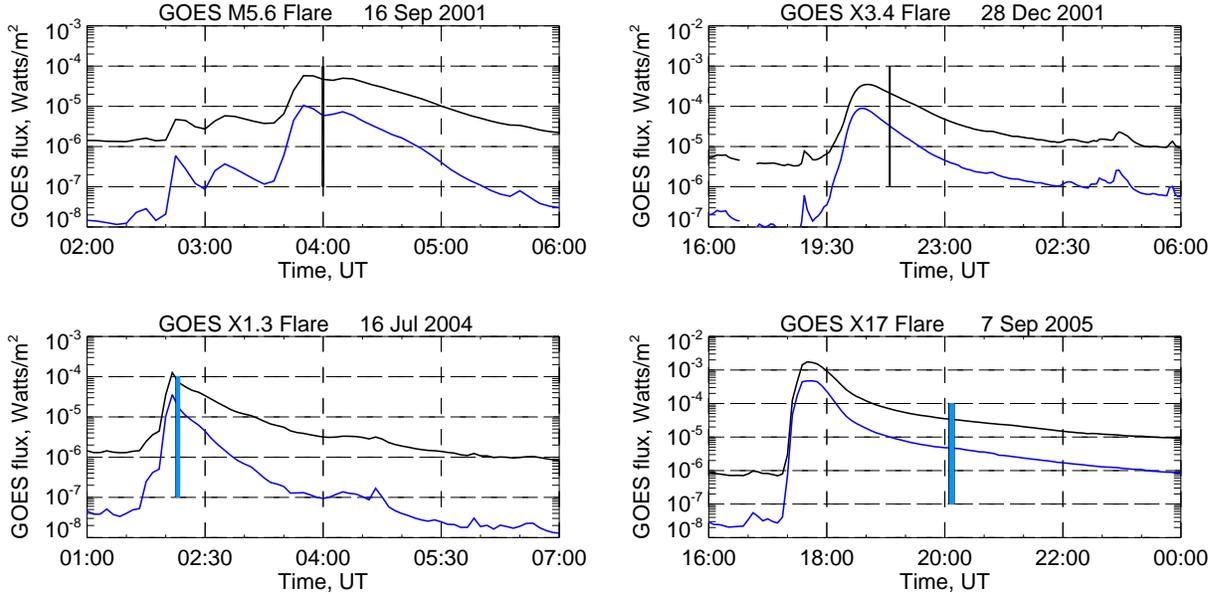}
    \caption{GOES X-ray lightcurves of the 4 flares. {The vertical lines}
    highlight the time when the SPIRIT spectroheliograms were obtained.}
    \label{GOES}
  \end{figure}

For the analysis, we have selected the following flares: M5.6 observed on 2001
September 16, X3.4 observed on 2001 December 28, X1.3 observed on 2004 July 16,
and X17 flare observed on 2005 September 7. All these flares are long duration
events (LDE), cover a broad range of flare intensity and have been registered in
different phases of their decay. The X-ray lightcurves of the flares measured by GOES
are shown in Figure~\ref{GOES}. Each SPIRIT spectroheliogram was obtained in a
single exposure (the exposures are denoted by vertical lines in the
Figure~\ref{GOES}). The exposure times for the M5.6 and X3.4 flares were 37
seconds, and 150 seconds for the X1.3 and X17 flares.  Some details of the
analyzed flares are given in the Table~\ref{Flares}.

  %\placefigure{GOES}

\begin{table}[!h]
    \caption{Flares class, peak time, active regions\label{Flares}}
    \scriptsize
    \begin{tabular}{|l|c|c|c|c|c|c|}    
      \tableline       
        GOES class  & Date  & GOES peak time, UT & SPIRIT obs. start, UT & NOAA AR & type, t$_{decay}$ \\
      \tableline
        M5.6  &  2001-09-16 & $\sim $ 03:50 & 03:59:36  & 9608  & LDE, 1 h 30 min\\
        X1.3  &  2004-07-16 & $\sim $ 02:05 & 02:07:54  & 10649 & LDE, $\sim 7$ h \\
        X3.4  &  2001-12-28 & $\sim $ 20:40 & 21:21:45  & 9767  & LDE, 2 h 40 min \\
        X17   &  2005-09-07 & $\sim $ 17:40 & 20:04:22  & 10808 & LDE, 5 h 50 min \\
      \tableline
    \end{tabular}
  \end{table}

\section{Data analysis}
\label{data_analysis}
\subsection{Interpretation of the SPIRIT spectroheliograms}
In a slitless scheme {a set of monochromatic solar images (each image in
a particular spectral line) is obtained on the detector}, shifted along the
dispersion axis. A small grazing incidence $\varphi \sim 1.5^{\circ}$ results in
a contraction of solar images along the dispersion axis. 

\begin{figure}[!ht] 
    \hspace{0.078\textwidth}
    \includegraphics[width = 0.5\textwidth]{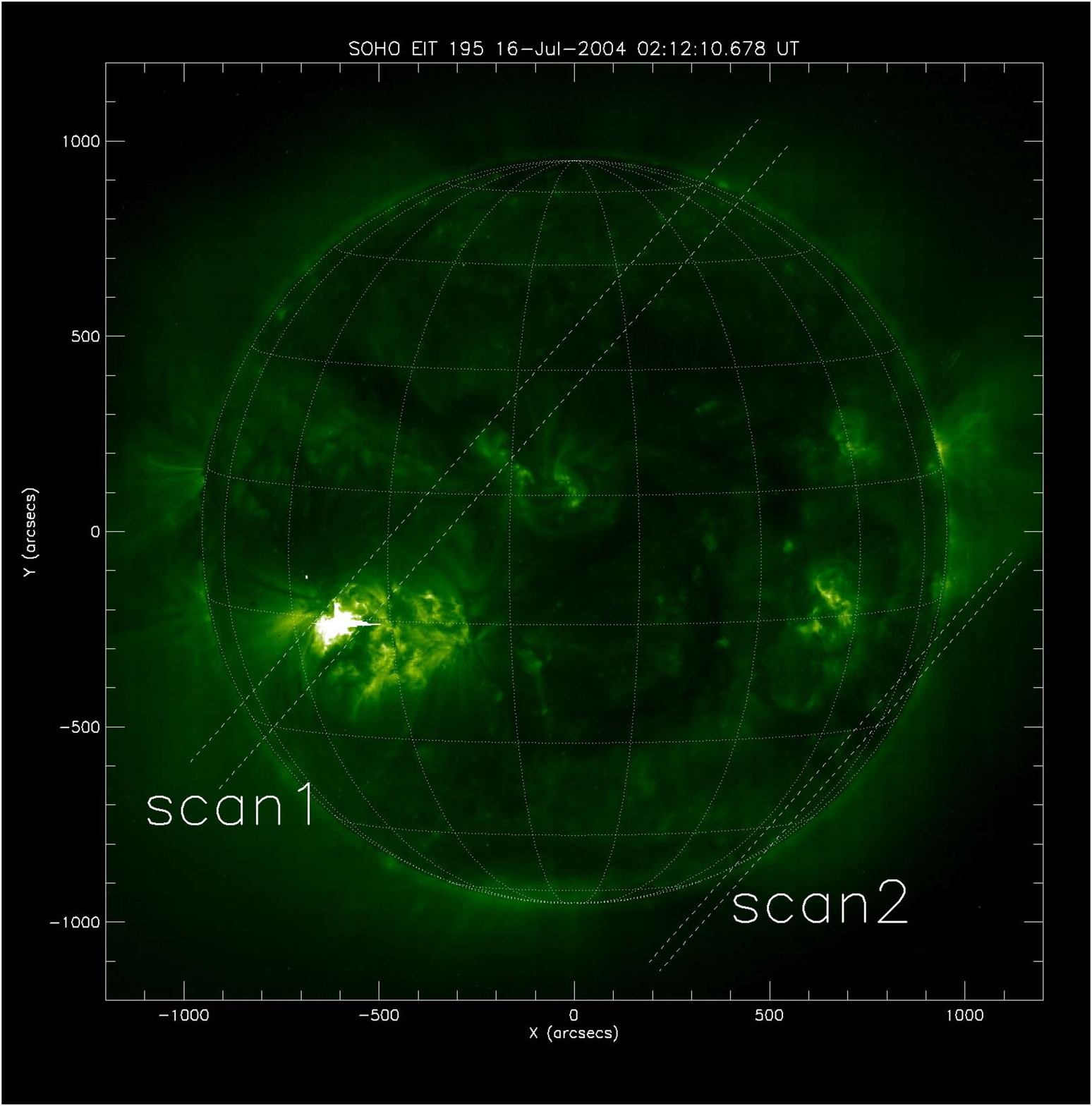} 
    \newline
    \includegraphics[width = \textwidth]{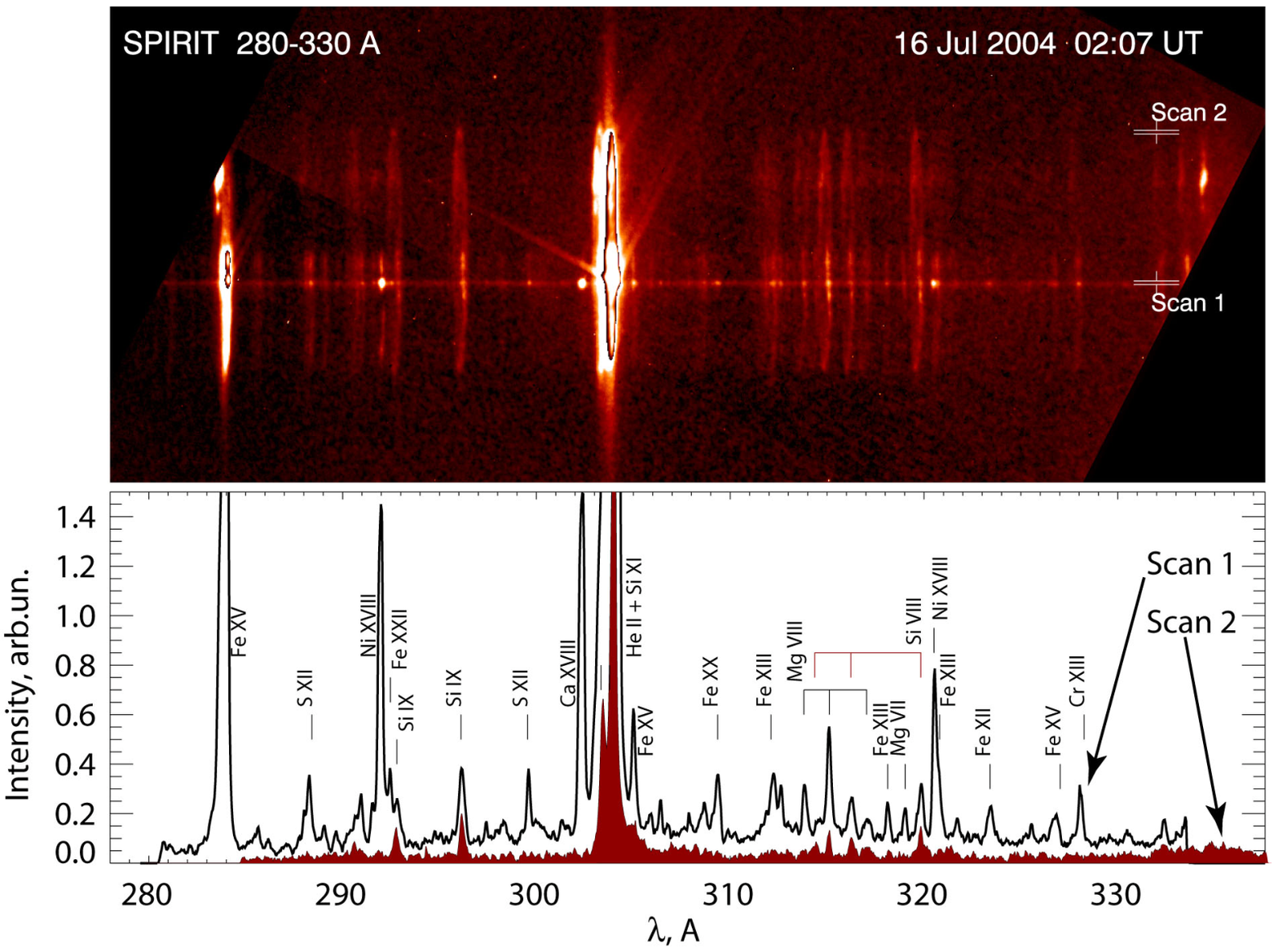}
    
    \caption{SPIRIT spectroheliogram for the 280--330~\AA\ range. Top panel:
    corresponding EIT 195 \AA\ image. Middle panel: the spectroheliogram
    registered on 2004 July 16; bottom panel: spectra of two regions ---
    flaring (scan1 --- X1.3 class flare) and arbitrary QS (scan2).  On the
    spectroheliogram each ellipse is a monochromatic solar disk image (in a
    particular spectral line) contracted along the dispersion axis.}
    \label{U304_overview}
  \end{figure}

\begin{figure}[!ht]
    \hspace{0.078\textwidth}
    \includegraphics[width = 0.5\textwidth]{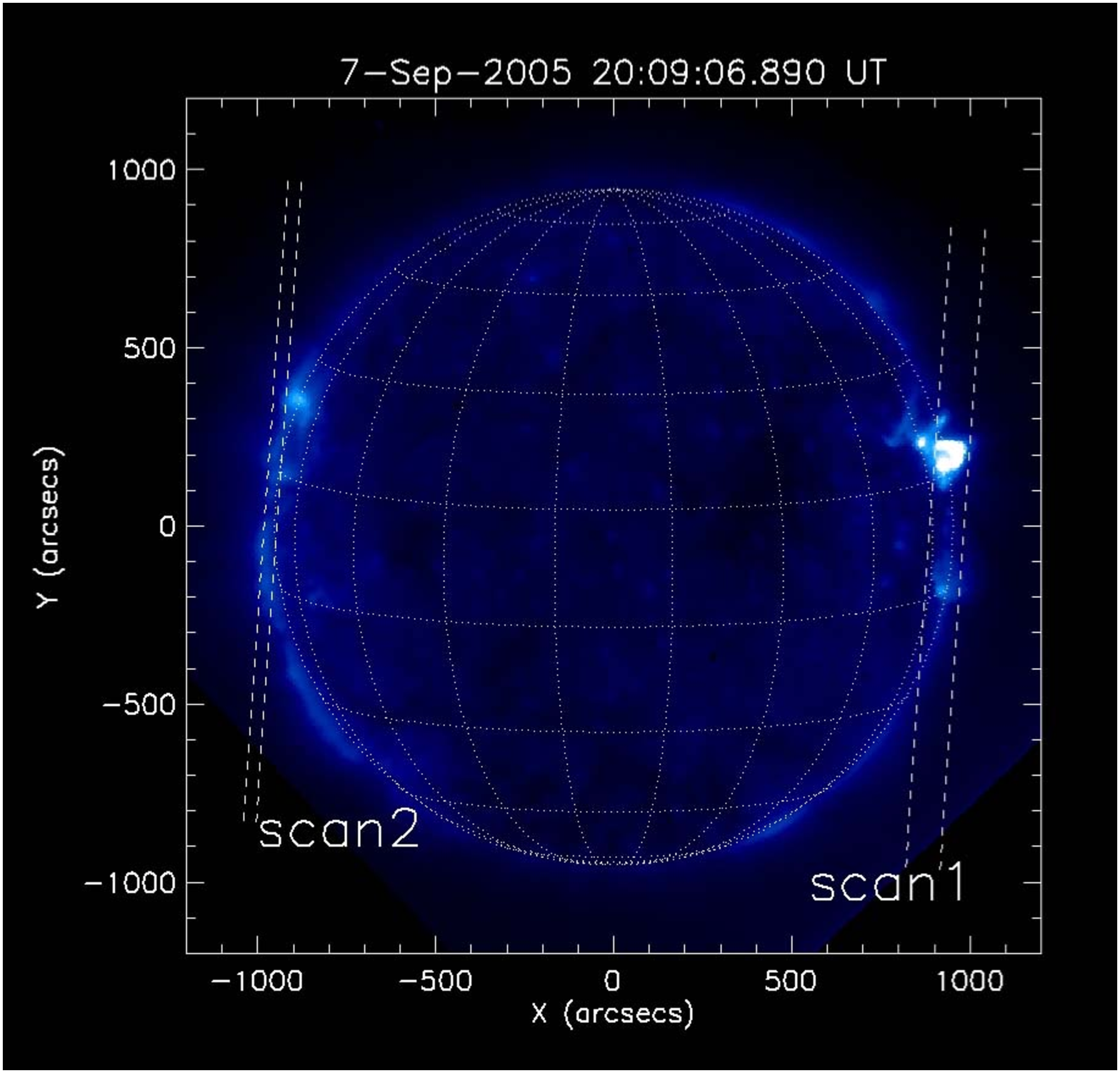} 
    \newline
    \includegraphics[width = \textwidth]{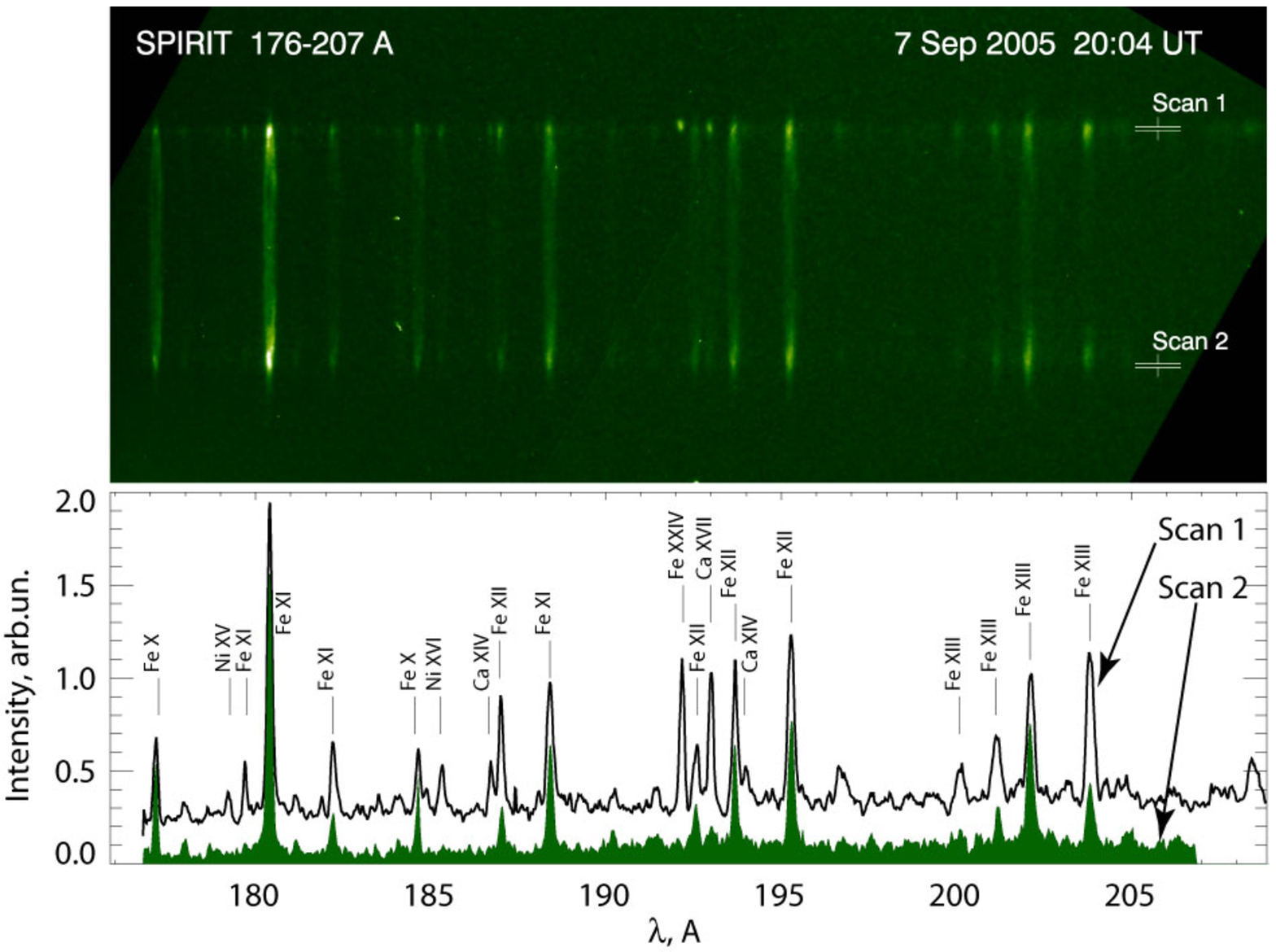}
    \vspace{-0.1\textwidth}
    \caption{SPIRIT spectroheliogram for the 176--207~\AA\ range. Top panel: corresponding SPIRIT 175 \AA\ telescope
    image (no EIT images were available on that day). Middle panel: the spectroheliogram registered on 2005 September 7; bottom panel:
    spectra of two regions --- flaring (scan1 --- X17 class flare) and arbitrary
    QS (scan2).  On the spectroheliogram each ellipse is a
    monochromatic solar disk image (in a particular spectral line) contracted
    along the dispersion axis.}
    \label{V190_overview}
  \end{figure}

Examples of the SPIRIT spectroheliograms are given in the {middle} panels of
Figure~\ref{U304_overview} and~\ref{V190_overview}. The X1.3 flare
(Figure~\ref{U304_overview}) appears as the bright horizontal line in the center
in the {\bf U304} channel. The X17 flare (Figure~\ref{V190_overview}) appears
as a brightening in the upper part of the solar disk in the {\bf V190} channel. On
the bottom panels of both Figures, directly extracted (``raw'') scans are given:
scan1 corresponds to the flare, and scan2 corresponds to arbitrary quite Sun
area. These raw scans are rows from respective images-arrays {with a roughly
assigned} linear wavelength scale. On the {top} panels of both Figures simultaneous
EUV images are given: EIT~195~\AA\ (Figure~\ref{V190_overview}) and
SPIRIT~175~\AA\ (Figure~\ref{U304_overview}; no simultaneous EIT image was
available). 

Comparison of the spectroheliograms and the extracted spectra shows that emission of
``cold'' coronal lines (like \ion{Si}{9}, \ion{Mg}{8}, \ion{Fe}{11},
\ion{Fe}{12} with $T_{max} \sim 1-2$~MK) originates from the whole solar disk,
 but due to contraction these monochromatic full-disk images look like ellipses.
 Emission of ``hot'' coronal lines (\ion{Ca}{17}, \ion{Ca}{18}, \ion{Fe}{20}, \ion{Fe}{22},
\ion{Fe}{24} with $T_{max} > 6$~MK ) is produced mainly in flaring regions, which
correspond to bright points in the spectroheliograms.

  %\placefigure{U304_overview}

  %\placefigure{V190_overview}

The interpretation of the spectroheliograms involves the following
steps: a) obtaining spectra of a particular region and determining the
wavelength scale; b) subtracting background from the spectra; c) identifying 
spectral lines with a subsequent analysis of spectral data. 

For obtaining spectra  from the spectroheliograms we have developed IDL
software, which implements a geometrical model of the spectroheliograph.
According to the model, for a particular point source the position on the
CCD-detector is calculated using its solar coordinates, wavelengths and several
parameters (such as direction to the solar center, {groove} density of the
diffraction grating, focal length and direction of the focusing mirror, relative
position of the CCD-detector etc.). The geometrical model automatically takes
into account contraction of solar disk images and non-linear wavelength scale
across the CCD-detector. Thus, to obtain spectra of a particular region and
calculate the wavelength scale for it, {one has only} to point to the region on the
solar disk.  The accuracy of the obtained wavelength scale is comparable to the
spectral size of 1 pixel ($\sim 0.04$~\AA).

For background subtraction we {used} a procedure similar to that of \citet{Thomas94}
--- we {interpolated} values outside spectral lines and {subtracted} the interpolation
from spectra.

Before the identification we also carefully {removed} strong \ion{Si}{11}
($\lambda=303.33$~\AA) and \ion{He}{2} (doublet $\lambda=303.78+.79$~\AA) blend
from spectra. This reveals the spectral lines of \ion{Ca}{18}, \ion{Ni}{14},
\ion{Fe}{15} ($\lambda \approx 302$~\AA) and \ion{Fe}{17}, \ion{Fe}{15}
($\lambda \approx 305$~\AA), which lie on the wings of the
\ion{Si}{11}/\ion{He}{2} blend. These lines are well distinguished on the
wings of the blend (see Figure~\ref{U304_overview}); therefore, we remove the
wings of the blend by interpolating the values outside the lines and manually
zero out the core of the blend.

In order to {identify the observational} spectrum and measure intensities of separate
spectral lines, we {produced} a synthetic spectrum,  which fits the observational data.
To produce a synthetic spectrum we use transitions and wavelengths from CHIANTI
(CHIANTI v.6 was used, \citet{Dere97, Dere2009}), set the line widths $\sigma$
in accordance with the instrument FWHM ($\sigma = -0.201+1.43 \cdot 10^{-3}
\cdot \lambda$~[\AA] for the {\bf V190} channel and $\sigma = 0.1$~[\AA] for the
{\bf U304} channel), and vary intensities to match the observational data.
However, straightforward fitting is not possible due to the relatively low spectral
resolution of SPIRIT --- $\sigma \sim 0.1$~\AA\ and blending of most of the
lines. 

\begin{figure}[!ht] 
    \includegraphics[scale = 1.0]{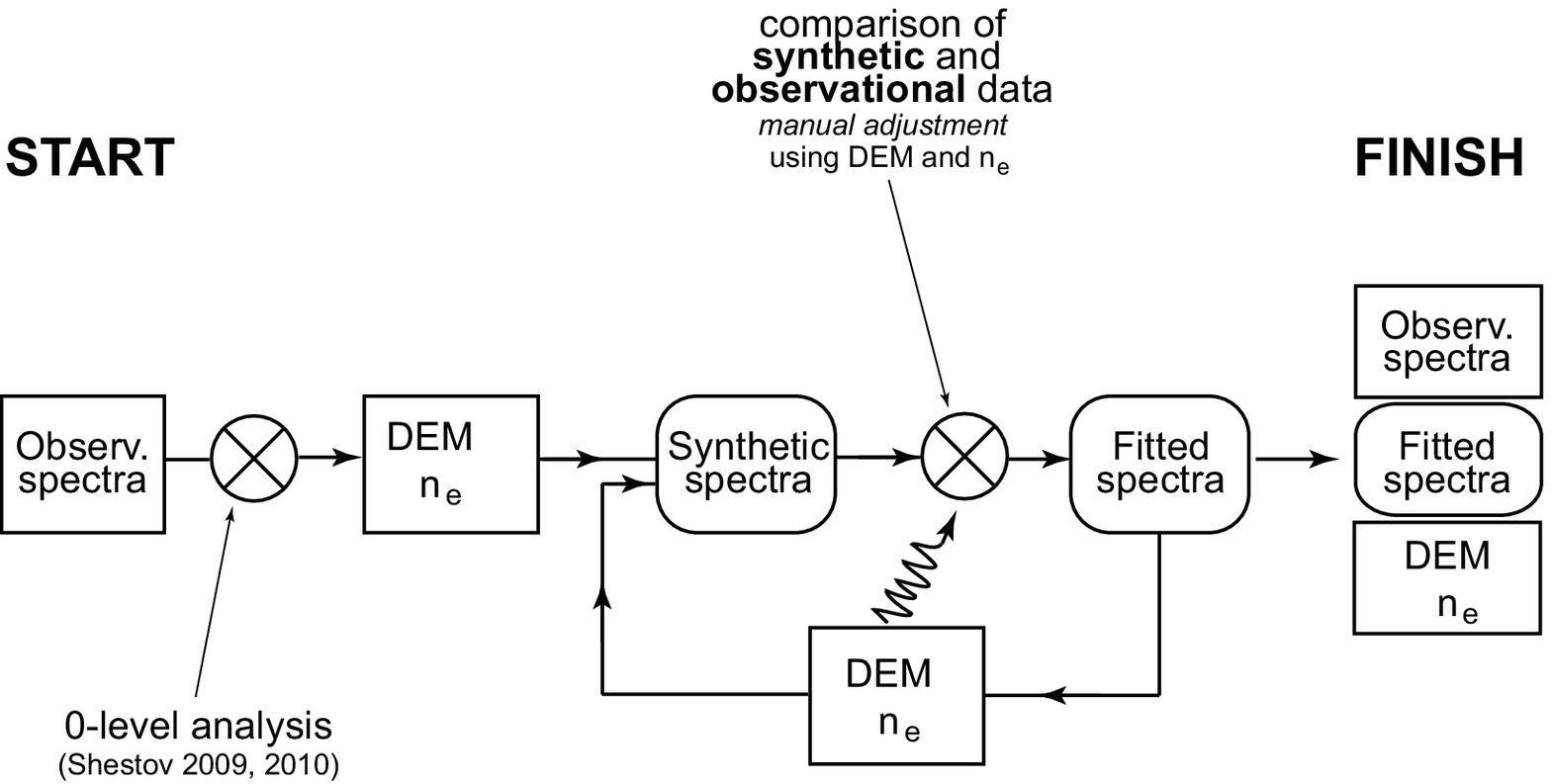}
    \caption{Procedure for spectra analysis.}
    \label{diagram}
  \end{figure}

We overcome this obstacle using an iterative procedure (see Figure~\ref{diagram}), which consists of
the initial step:  
\begin{itemize}
    \item Measurement of
intensities of a small number of spectral lines and calculation of plasma
parameters --- DEM and $n_e$ (see \citet{Shestov2009e,Shestov2010e});
\end{itemize}
and further (iterative) steps:
\begin{itemize}
    \item Calculation of the synthetic spectra;
    \item Automated adjustment of spectral line intensities to match the
	    observational data. During the adjustment, the ratio of the blended
	    lines is kept constant;
    \item Manual adjustment of intensities of particular spectral lines. Using
	    the DEM and $n_e$ analysis data we adjust intensities of
	    blended lines {to reach a better} agreement with theory (reducing
	    $\chi^2$ in DEM reconstruction and compliance with other lines in
	    L-function analysis --- see below); 
    \item Calculation of DEM and $n_e$.
\end{itemize}
The larger number of spectral lines, used for analysis during iterative steps,
almost completely eliminates errors due to possible misidentification or other
errors. The iterative procedure turned out to be fast and stable --- after the
second step there are no considerable changes in DEMs and synthetic
spectra. So, in our approach plasma diagnostic was an essential part of
the line identification --- we used plasma parameters to resolve blended lines. 

  %\placefigure{U304_overview}

For the calculation of synthetic spectra we used standard CHIANTI procedures
\texttt{ch\_synthetic} and \texttt{make\_chianti\_spec}, coronal abundances \texttt{sun\_coronal.abund} and
\texttt{mazzotta\_etal.ioneq} ionization equilibrium.

For the DEM reconstruction we used a Genetic Algorithm (GA) \citep{Siarkowski2008}.
The algorithm is based on ideas of biological evolution and natural selection.
It starts from randomly chosen initial populations of different DEMs and produces a new
generation of DEMs by crossover and mutations. The procedure stops
when a local $\chi ^2$ minimum is found. The peculiar feature of the method is
that since it is based on a random evolution, different runs of the procedure on
a single data set give different (but similar) results. The discrepancy
among different runs directly shows the confidence of the DEM reconstruction.  

For the DEM analysis we carefully chose 46 spectral lines (Table~\ref{DEMlines})
--- almost all strong spectral lines, observed by SPIRIT. The exceptions are
\ion{Fe}{15} 284.16~\AA\ and the \ion{Si}{11}/\ion{He}{2} blend with $\lambda
\sim 304$~\AA. Both these lines are very intense, which is likely to cause
saturation of the SPIRIT detector. Also, the observed intensity of the
\ion{Fe}{15} line shows systematic discrepancy with other \ion{Fe}{15} lines
(we will discuss possible reasons later).  Nevertheless, the spectral
lines analysed cover a wide temperature range --- from $T_{max} \sim 1$~MK
(\ion{Mg}{8}) to $T_{max} \sim 16$~MK (\ion{Fe}{24}). Using of a large number of
lines almost completely {eliminates the sensitivity} of the reconstructed DEM to the
intensity of a particular line, improving reliability of the reconstruction. 

 \begin{table}[!h]
    \caption{List of spectral lines used in the DEM reconstruction.\label{DEMlines}}
    \scriptsize
    \begin{tabular}{|r|l|l|c||r|l|l|c||r|l|l|c|}    
      \tableline       
        N & Ion & $\lambda$, \AA\ & $\log T_m$, K & N & Ion & $\lambda$, \AA\ & $\log T_m$, K & N & Ion & $\lambda$, \AA\ & $\log T_m$, K \\
      \tableline
       1 & Fe XI     & 180.41 & 6.2 &   17 & Fe XIII   & 196.54 & 6.3 &   33 & Fe XIII   & 312.11 & 6.3 \\ 
 2 & Fe XI     & 182.17 & 6.2 &   18 & Fe XII    & 196.64 & 6.3 &   34 & Fe XII    & 312.25 & 6.3 \\
 3 & Fe X      & 184.54 & 6.2 &   19 & Fe XIII   & 200.02 & 6.3 &   35 & Mg VIII   & 313.74 & 6.0 \\
 4 & Ni XVI    & 185.23 & 6.4 &   20 & Fe XIII   & 202.04 & 6.3 &   36 & Si VIII   & 314.36 & 6.1 \\
 5 & Ca XIV    & 186.61 & 6.6 &   21 & Fe XIII   & 203.83 & 6.3 &   37 & Mg VIII   & 315.02 & 6.0 \\
 6 & Fe XII    & 186.89 & 6.3 &   22 & S XI      & 285.82 & 6.3 &   38 & Si VIII   & 316.22 & 6.1 \\
 7 & Fe XXI    & 187.93 & 7.1 &   23 & Ni XVI    & 288.17 & 6.4 &   39 & Mg VIII   & 317.03 & 6.0 \\
 8 & Fe XI     & 188.23 & 6.2 &   24 & Ni XVIII  & 291.98 & 6.8 &   40 & Fe XIII   & 318.13 & 6.3 \\
 9 & Fe XI     & 188.30 & 6.2 &   25 & Fe XXII   & 292.46 & 7.1 &   41 & Mg VII    & 319.03 & 5.8 \\
10 & Fe XXIV   & 192.03 & 7.2 &   26 & Si IX     & 292.76 & 6.2 &   42 & Si VIII   & 319.84 & 6.1 \\
11 & Fe XII    & 192.39 & 6.3 &   27 & Si IX     & 296.11 & 6.2 &   43 & Ni XVIII  & 320.57 & 6.8 \\
12 & Fe XI     & 192.83 & 6.2 &   28 & S XII     & 299.54 & 6.3 &   44 & Fe XIII   & 320.81 & 6.3 \\
13 & Ca XVII   & 192.85 & 6.8 &   29 & Ca XVIII  & 302.19 & 7.0 &   45 & Fe XII    & 323.41 & 6.3 \\
14 & Fe XII    & 193.51 & 6.3 &   30 & Fe XV     & 302.33 & 6.3 &   46 & Fe XVII   & 323.65 & 6.8 \\
15 & Ca XIV    & 193.87 & 6.6 &   31 & Fe XV     & 304.89 & 6.3 &   47 & Fe XV     & 327.03 & 6.3 \\
16 & Fe XII    & 195.12 & 6.3 &   32 & Fe XX     & 309.29 & 7.0 &      &           &        &     \\

      \tableline
    \end{tabular}
  \end{table}

  Electron density $n_e$ was obtained using a modified L-function
analysis~\citep{Landi98}. According to the original method proposed by the
authors, L-functions of all spectral lines of a particular ion should intersect
at a single point, corresponding to the density of the emitting plasma.  The
L-function of a spectral line is defined as a ratio of measured intensity over
the contribution function $G(T_o,n_e)$, plotted as a function of density. We
slightly simplify the definition of L-function by using $T_{max}$ instead of
$T_o$ (specially computed temperature) and plot the L-functions for major lines
of the \ion{Fe}{11}, \ion{Fe}{12}, \ion{Fe}{13}, \ion{Fe}{15}, \ion{Mg}{8}, and
\ion{Ni}{16} ions. 

\subsection{Absolute calibration of SPIRIT fluxes}
\label{absolute_calibration}
No absolute ground calibration was carried out before the launch of the SPIRIT.
Lack of calibration cripples spectroheliograph diagnostic capabilities. However,
the spectral ranges of the SPIRIT {\bf V190} and {\bf U304} channels overlap with
spectral responses of the EIT 195 \AA\ and 304 \AA\ channels, and it is possible
to cross-calibrate SPIRIT data with EIT data.
  
The total flux $F$ in an EIT image expressed in [dn] (digital numbers) can be expressed as: 
\begin{equation}
F=\int s(\lambda) b(\lambda) d \lambda,
\label{E:EIT_flux}
\end{equation}
where $s(\lambda)$ --- is real incident spectral flux, units [erg/s/cm$^2$/\AA], $b(\lambda)$
--- is the EIT spectral sensitivity, expressed in units [cm$^2$ dn/erg] and obtained
with the \texttt{eit\_parm} function from Solar Software. The $s(\lambda)$ can
be expressed as:
\begin{equation}
s(\lambda)=k \cdot i(\lambda),
\label{E:Real flux}
\end{equation}
where $i(\lambda)$ [DN] --- is
spectral flux measured by SPIRIT and $k$ [erg/s/cm$^2$/\AA/DN] --- is the calibration coefficient to be found. From equations (\ref{E:EIT_flux}) and (\ref{E:Real flux}) we calculate $k$:
\begin{equation}
k=\frac{F}{\int i(\lambda) b(\lambda) d \lambda}
\end{equation}

\begin{figure} 
    \plotone{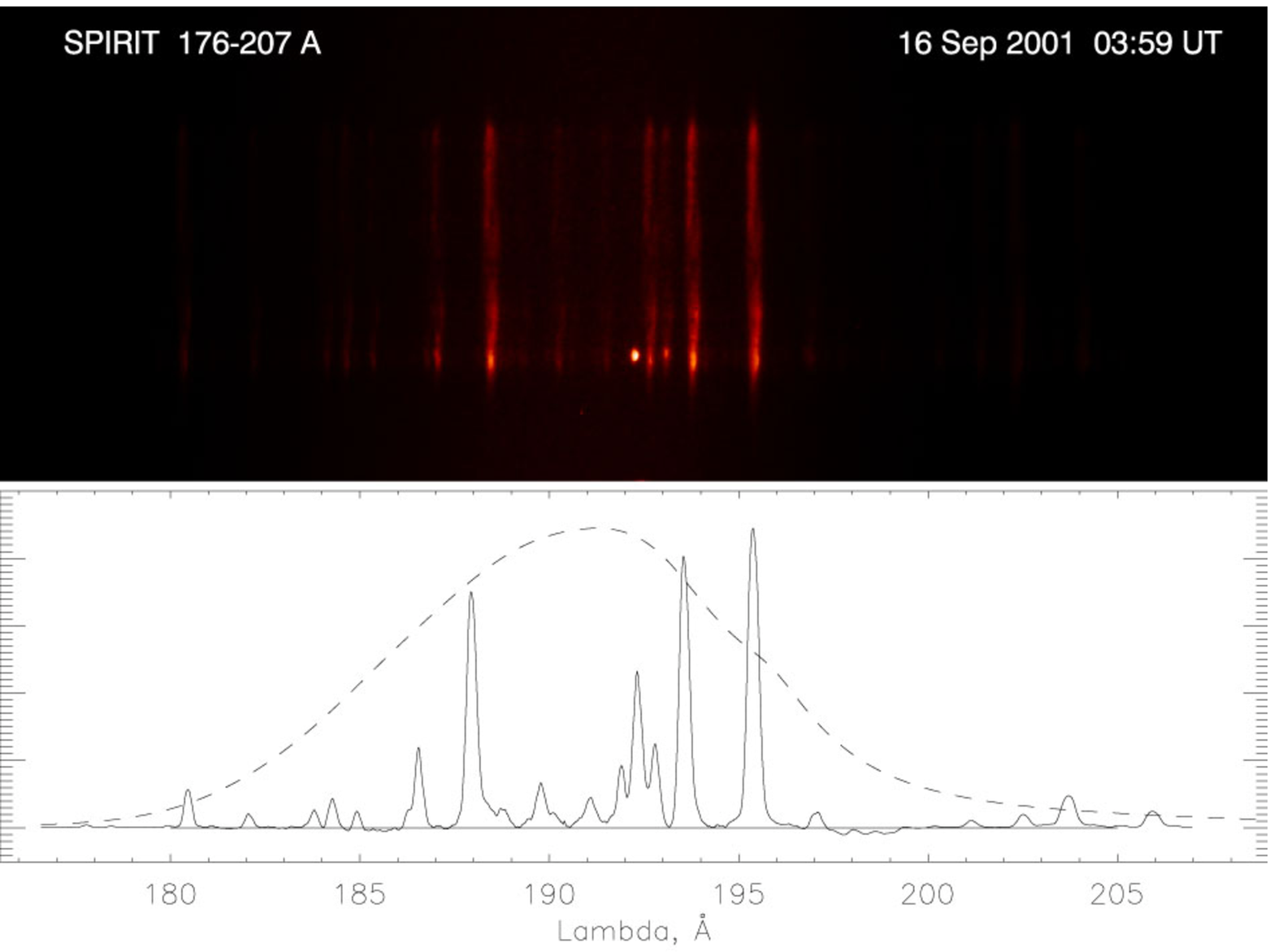}
    \caption{Top panel: the {\bf V190} spectoheliogram registered on 2001 Sep 9,
    convolved with EIT 195~\AA\ bandpass function. Bottom panel: solid line ---
    SPIRIT spectra from the forementioned spectroheliogram integrated along the
    spatial axis and multiplied by EIT 195~\AA\ bandpass function. Dashed line
    --- normalized EIT 195~\AA\ bandpass function.}
    \label{V190_calibr}
  \end{figure}

The relative spectral flux $i(\lambda)$ was obtained by integrating the whole
SPIRIT spectroheliogram along the spatial axis. The total EIT flux $F$ was
obtained by integrating the whole EIT image (195 \AA\ for the {\bf V190}
channel, and 304 \AA\ for the {\bf U304} channel). The {\bf V190} channel
spectroheliogram containing the M5.6 flare and the whole-Sun relative spectral
flux $i(\lambda)$, both multiplied by $b(\lambda)$, is given 
in Figure~\ref{V190_calibr}.

We carried out this procedure for all flare spectra presented in this work and
converted the spectra into physical units.  However, we believe that calibration
coefficient $k$ obtained for the {\bf U304} channel is less reliable than for
the {\bf V190} channel, due to possible nonlinear response of the SPIRIT
detector to the intense fluxes. That is why we performed an independent
verification of the obtained absolute fluxes. The verification uses a
spectroscopic approach and consists of the following: during the DEM
calculation the $\chi^2$ parameter is minimized. We introduced calibration
correction factor $\kappa$ for the {\bf U304} channel and calculated $\chi^2$
values for a range of $\kappa$ values. The minimum $\chi^2$ value gives best
cross-calibration $\kappa$ from the spectroscopic point of view. The calculated
best $\kappa$ values are $1.0$, $0.63$, $1.1$ for the M5.6, X1.3 and
X17 flares.  These $\kappa$ have been taken into account --- we modified data
in {\bf U304} channel spectra.

\section{Results}
We have analysed spectra of the four flares and note three main results of our analysis:
\begin{itemize}
	\item{A catalog of EUV spectral lines observed in large solar flares;}
	\item{DEM and $n_e$ of the emitting plasma;}
	\item{A benchmark of the atomic database, by analysing ratios
		of the observed and calculated spectral line intensities.}
\end{itemize}

\subsection{Catalog of spectral lines} 
\begin{figure} 
  \plotone{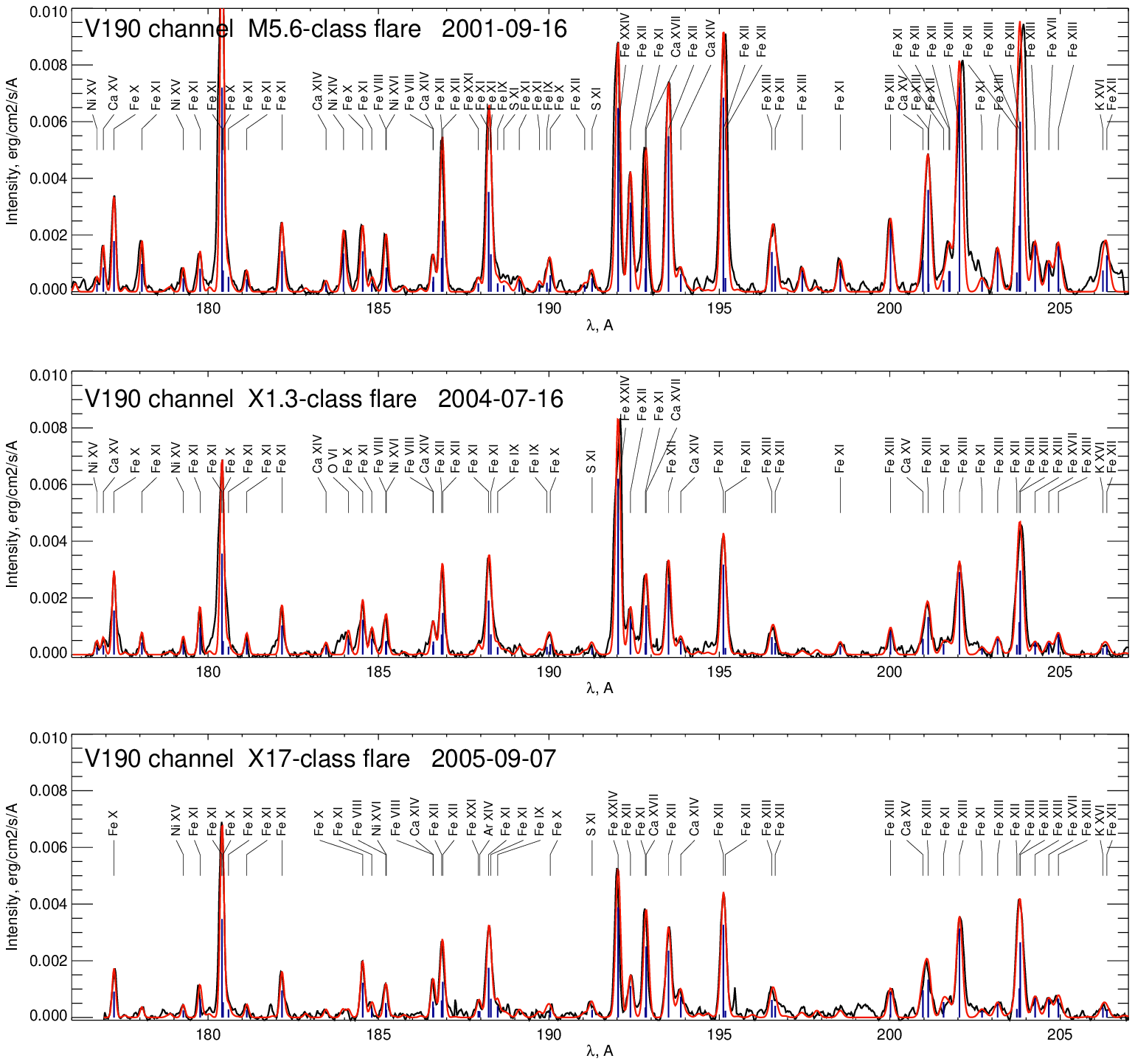}
  \caption{{\bf V190} spectra of flares: X1.3 observed on 2004-07-16; M5.6
  observed on 2001-09-16; and X17 observed on 2005-09-07. The black curve denotes
  observational data, blue vertical lines denote individual spectral lines from
  catalog, and the red curve denotes fitted spectra (calculated using instrument
  FWHM).}
  \label{V190_flares}
\end{figure}

\begin{figure} 
  \includegraphics[scale=0.95]{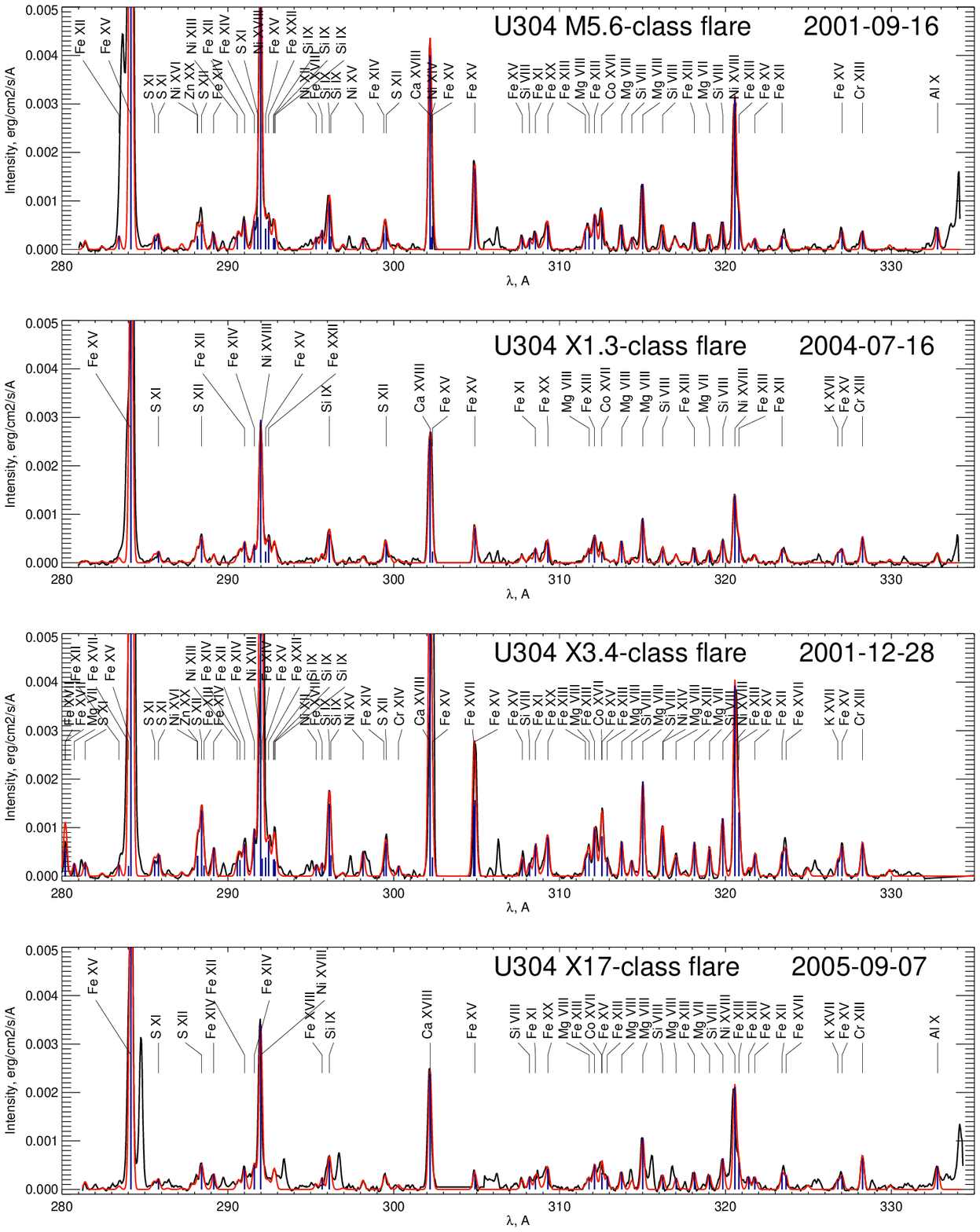}
  \caption{{\bf U304} spectra of flares: X1.3 observed on 2004-07-16; M5.6
  observed on 2001-09-16; X17 observed on 2005-09-07; and X3.4 observed on
  2001-12-28. The black curve denotes observational data, blue vertical
  lines denote individual spectral lines from catalog, and the red curve
  denotes fitted spectra (calculated using instrument FWHM).}
  \label{U304_flares}
\end{figure}

Comparisons of observational and fitted spectra are given in
Figure~\ref{V190_flares} ({\bf V190}) and Figure~\ref{U304_flares} ({\bf U304}).
The black curve denotes observational data, blue vertical lines denote
individual spectral lines from the catalog, and the red curve denotes fitted spectra.

The catalog of spectral lines is given in Table~\ref{V190_table} ({\bf V190}
channel) and Table~\ref{U304_table} ({\bf U304} channel). Only the strongest 70
lines were included in the tables, but during the identification we operated
with a larger number of lines.

%\placefigure{V190_flares}

%\placefigure{U304_flares}

\begin{deluxetable}{rccccccc|rccccccc}
  \tabletypesize{\scriptsize}	
  \rotate
  \tablecaption{Intensities of spectral lines [$\cdot 10^{-5}$ erg/s/cm$^2$] in the
	  solar flares in the {\bf V190} channel.\label{V190_table}}
  \tablewidth{0pt}
  \tablehead{   &               &                          &                          &  \multicolumn{4}{c}{Intensity} &
                &               &                          &                          &  \multicolumn{4}{c}{Intensity} \\
    \colhead{N} \vspace{-0.2cm} & \colhead{Ion} & \colhead{$\lambda$,~\AA} & \colhead{$T_{max}$,~K} &    &    &    &   &
    \colhead{N}                 & \colhead{Ion} & \colhead{$\lambda$,~\AA} & \colhead{$T_{max}$,~K} &    &    &    &   \\  
           &   &   &   &    \colhead{M5.6} & \colhead{X1.3} & \colhead{X3.4} & \colhead{X17} &
           &   &   &   &    \colhead{M5.6} & \colhead{X1.3} & \colhead{X3.4} & \colhead{X17}  }
  \startdata
        1 &    Ni XV & 176.10 & 6.4 &   3.6 &  ---  &  N/A  &  ---  & 39 &     S XI & 191.27 & 6.3 &  11.9 &   5.9 &  N/A  &   6.7 \\
      2 &    Ni XV & 176.74 & 6.4 &   6.8 &   6.4 &  N/A  &  ---  & 40 &  Fe XXIV & 192.03 & 7.2 & 162.0 & 155.0 &  N/A  &  97.1 \\
      3 &    Ca XV & 176.93 & 6.7 &  21.3 &   8.0 &  N/A  &  ---  & 41 &   Fe XII & 192.39 & 6.3 &  78.5 &  31.5 &  N/A  &  27.6 \\
      4 &     Fe X & 177.24 & 6.1 &  44.6 &  38.8 &  N/A  &  22.7 & 42 &    Fe XI & 192.83 & 6.2 &  20.5 &  10.1 &  N/A  &   8.7 \\
      5 &    Fe XI & 178.06 & 6.2 &  24.3 &  10.6 &  N/A  &   4.9 & 43 &  Ca XVII & 192.85 & 6.8 &  74.1 &  43.4 &  N/A  &  62.7 \\
      6 &    Ni XV & 179.27 & 6.4 &  11.8 &   8.9 &  N/A  &   6.2 & 44 &   Fe XII & 193.51 & 6.3 & 137.0 &  61.9 &  N/A  &  58.9 \\
      7 &    Fe XI & 179.76 & 6.2 &  20.2 &  23.9 &  N/A  &  16.2 & 45 &    Fe XI & 193.51 & 6.2 &   3.5 &  ---  &  N/A  &  ---  \\
      8 &    Fe XI & 180.41 & 6.2 & 180.0 &  89.0 &  N/A  &  86.8 & 46 &     Fe X & 193.71 & 6.1 &   3.4 &  ---  &  N/A  &   3.2 \\
      9 &     Fe X & 180.44 & 6.1 &  18.7 &  11.9 &  N/A  &  13.4 & 47 &   Ca XIV & 193.87 & 6.6 &  15.4 &  11.8 &  N/A  &  18.2 \\
     10 &    Fe XI & 180.60 & 6.2 &  12.9 &   6.9 &  N/A  &   6.9 & 48 &   Ni XVI & 194.02 & 6.4 &   3.3 &  ---  &  N/A  &  ---  \\
     11 &    Fe XI & 181.14 & 6.2 &  10.9 &  11.0 &  N/A  &   6.7 & 49 &   Fe XII & 195.12 & 6.3 & 171.0 &  79.2 &  N/A  &  81.7 \\
     12 &    Fe XI & 182.17 & 6.2 &  35.7 &  25.6 &  N/A  &  23.7 & 50 &   Fe XII & 195.18 & 6.3 &  12.1 &   5.7 &  N/A  &   5.8 \\
     13 &   Ca XIV & 183.46 & 6.6 &   6.1 &   6.6 &  N/A  &   3.7 & 51 &  Fe XIII & 196.54 & 6.3 &  35.0 &  15.3 &  N/A  &  15.5 \\
     14 &   Ni XIV & 183.97 & 6.3 &  33.6 &  ---  &  N/A  &  ---  & 52 &   Fe XII & 196.64 & 6.3 &  22.6 &  10.2 &  N/A  &  10.3 \\
     15 &     O VI & 184.12 & 6.3 &   3.4 &  13.4 &  N/A  &   4.4 & 53 &  Fe XIII & 197.43 & 6.3 &  17.3 &  ---  &  N/A  &  ---  \\
     16 &     Fe X & 184.54 & 6.1 &  35.6 &  30.7 &  N/A  &  30.6 & 54 &    Fe IX & 197.86 & 6.1 &   3.8 &   3.4 &  N/A  &   5.0 \\
     17 &    Fe XI & 184.80 & 6.2 &  11.8 &  13.8 &  N/A  &   8.0 & 55 &    Fe XI & 198.55 & 6.2 &   6.9 &   3.7 &  N/A  &   3.7 \\
     18 &  Fe VIII & 185.21 & 5.8 &  11.5 &  11.3 &  N/A  &  12.6 & 56 &  Fe XIII & 200.02 & 6.3 &  55.2 &  20.4 &  N/A  &  21.9 \\
     19 &   Ni XVI & 185.23 & 6.4 &  21.3 &  12.0 &  N/A  &   6.6 & 57 &    Ca XV & 200.97 & 6.7 &  27.8 &  13.8 &  N/A  &  23.9 \\
     20 &  Fe VIII & 186.60 & 5.7 &   8.9 &   8.1 &  N/A  &  14.0 & 58 &    Fe XX & 201.05 & 7.0 &   3.8 &  ---  &  N/A  &   3.5 \\
     21 &   Ca XIV & 186.61 & 6.6 &  13.0 &  11.5 &  N/A  &   8.7 & 59 &  Fe XIII & 201.13 & 6.3 &  89.8 &  33.1 &  N/A  &  33.3 \\
     22 &   Fe XII & 186.85 & 6.3 &  29.7 &  17.6 &  N/A  &  15.0 & 60 &   Fe XII & 201.14 & 6.3 &   7.1 &  ---  &  N/A  &  ---  \\
     23 &   Fe XII & 186.89 & 6.3 &  62.5 &  36.6 &  N/A  &  31.4 & 61 &    Fe XI & 201.58 & 6.2 &  35.1 &  13.6 &  N/A  &  13.5 \\
     24 &   Fe XXI & 187.93 & 7.1 &   7.3 &   4.9 &  N/A  &   5.6 & 62 &   Fe XII & 201.74 & 6.3 &  13.0 &   4.5 &  N/A  &   4.4 \\
     25 &   Ar XIV & 187.97 & 6.6 &  ---  &  ---  &  N/A  &   5.5 & 63 &   Fe XII & 201.76 & 6.3 &  10.0 &   3.4 &  N/A  &   3.4 \\
     26 &   Fe XII & 188.17 & 6.3 &   3.4 &  ---  &  N/A  &  ---  & 64 &  Fe XIII & 202.04 & 6.3 & 180.0 &  72.7 &  N/A  &  78.4 \\
     27 &    Fe XI & 188.23 & 6.2 &  88.0 &  47.5 &  N/A  &  43.8 & 65 &    Fe XI & 202.71 & 6.2 &   9.9 &   5.3 &  N/A  &   5.3 \\
     28 &    Fe XI & 188.30 & 6.2 &  33.0 &  17.8 &  N/A  &  16.4 & 66 &  Fe XIII & 203.16 & 6.3 &  34.7 &  14.0 &  N/A  &  12.0 \\
     29 &    Fe IX & 188.50 & 6.0 &   7.3 &   6.7 &  N/A  &  10.0 & 67 &   Fe XII & 203.73 & 6.3 &  16.8 &   8.5 &  N/A  &   7.5 \\
     30 &     S XI & 188.68 & 6.3 &   5.5 &  ---  &  N/A  &  ---  & 68 &  Fe XIII & 203.80 & 6.3 &  58.1 &  28.5 &  N/A  &  25.6 \\
     31 &    Ar XI & 188.81 & 6.3 &   3.7 &  ---  &  N/A  &  ---  & 69 &  Fe XIII & 203.83 & 6.3 & 150.0 &  74.1 &  N/A  &  66.2 \\
     32 &    Fe XI & 189.13 & 6.2 &   8.6 &   4.6 &  N/A  &   4.6 & 70 &  Fe XIII & 204.26 & 6.3 &  40.5 &  10.6 &  N/A  &  17.1 \\
     33 &    Fe XI & 189.72 & 6.2 &   6.6 &   3.6 &  N/A  &   3.5 & 71 &  Fe XVII & 204.67 & 6.9 &  24.7 &  10.5 &  N/A  &  15.8 \\
     34 &    Fe IX & 189.94 & 6.1 &   7.9 &   6.7 &  N/A  &   5.0 & 72 &  Fe XIII & 204.95 & 6.3 &  39.8 &  17.7 &  N/A  &  16.4 \\
     35 &     Fe X & 190.04 & 6.2 &  14.4 &   9.2 &  N/A  &   5.1 & 73 &    K XVI & 206.25 & 6.7 &  18.6 &   5.5 &  N/A  &   8.6 \\
     36 &   Fe XII & 190.07 & 6.3 &   4.0 &  ---  &  N/A  &  ---  & 74 &   Fe XII & 206.37 & 6.3 &  32.0 &   7.2 &  N/A  &   6.5 \\
     37 &   Fe XII & 191.05 & 6.3 &   5.4 &  ---  &  N/A  &  ---  & 75 &   Ca XVI & 208.60 & 6.7 &  ---  &  ---  &  N/A  &  13.1 \\
     38 &    Fe IX & 191.22 & 6.0 &  ---  &  ---  &  N/A  &   4.1 &    &          &        &     &       &       &       &       \\

  \enddata
  \tablecomments{Columns correspond to different
  flares, rows denote different spectral lines. The flux in a particular spectral
  line corresponds to the whole flaring region (and do not contain $\frac{1}{sr}$ factor).
	  Minus '--' sign denotes that the line was too weak in a particular spectrum,
	  'N/A' in the X3.4 flare means that the flare was not observed by {\bf V190}
	  channel of SPIRIT.}
  \end{deluxetable}

\begin{deluxetable}{rccccccc|rccccccc}
  \tabletypesize{\scriptsize}
  \rotate
  \tablecaption{Intensities of spectral lines [$\cdot 10^{-5}$ erg/s/cm$^2$] in the
	  {\bf U304} channel.\label{U304_table}}
  \tablewidth{0pt}
  \tablehead{
                &               &                          &                          &  \multicolumn{4}{c}{Intensity} &
                &               &                          &                          &  \multicolumn{4}{c}{Intensity} \\
    \colhead{N} \vspace{-0.2cm} & \colhead{Ion} & \colhead{$\lambda$,~\AA} & \colhead{$T_{max}$,~K} &    &    &    &   &
    \colhead{N}                 & \colhead{Ion} & \colhead{$\lambda$,~\AA} & \colhead{$T_{max}$,~K} &    &    &    &   \\  
           &   &   &   &    \colhead{M5.6} & \colhead{X1.3} & \colhead{X3.4} & \colhead{X17} &
           &   &   &   &    \colhead{M5.6} & \colhead{X1.3} & \colhead{X3.4} & \colhead{X17}  }
  \startdata
            1 &  Fe XVII & 280.20 & 6.8 &  11.4 &  ---  &  14.2 &  ---  & 39 &  Fe XVII & 304.82 & 6.8 &   3.9 &  ---  &  36.2 &  ---  \\ 
      2 &   Mg VII & 280.74 & 5.8 &  ---  &  ---  &   6.5 &  ---  & 40 &    Fe XV & 304.89 & 6.3 &  41.4 &  17.6 &  39.1 &   7.6 \\
      3 &     S XI & 281.40 & 6.3 &   4.5 &  ---  &   7.0 &   4.0 & 41 &    Fe XV & 307.75 & 6.3 &   6.5 &   3.6 &   8.5 &   4.3 \\
      4 &   Fe XII & 283.44 & 6.3 &   7.0 &  ---  &   5.8 &  ---  & 42 &  Si VIII & 308.19 & 5.9 &   5.6 &   3.4 &   6.7 &   6.3 \\
      5 &  Fe XVII & 284.01 & 6.8 &  ---  &  ---  &   5.2 &  ---  & 43 &    Fe XI & 308.55 & 6.2 &   8.5 &   7.4 &  16.0 &   5.3 \\
      6 &    Fe XV & 284.16 & 6.3 & 354.0 & 204.0 & 610.0 & 202.0 & 44 &  Fe XIII & 308.69 & 6.3 &  ---  &  ---  &  ---  &   4.8 \\
      7 &     S XI & 285.59 & 6.3 &   6.6 &   3.7 &   8.0 &   3.1 & 45 &   Ni XVI & 309.18 & 6.4 &   3.8 &  ---  &  ---  &  ---  \\
      8 &     S XI & 285.82 & 6.3 &   7.7 &   5.6 &  10.9 &   5.3 & 46 &    Fe XX & 309.29 & 7.0 &  12.4 &  10.2 &  19.8 &  10.5 \\
      9 &   Fe XIV & 287.87 & 6.3 &   4.1 &  ---  &  ---  &   3.1 & 47 &  Fe XIII & 311.55 & 6.3 &   9.4 &  ---  &   8.9 &  ---  \\
     10 &   Ni XVI & 288.17 & 6.4 &   6.9 &   3.9 &  10.4 &  ---  & 48 &  Mg VIII & 311.77 & 5.9 &   8.6 &   6.4 &  11.8 &   5.5 \\
     11 &    Zn XX & 288.18 & 6.8 &   5.4 &   3.9 &  10.4 &   3.0 & 49 &  Fe XIII & 312.11 & 6.3 &  17.6 &  13.1 &  24.2 &  12.5 \\
     12 &    S XII & 288.42 & 6.4 &  13.0 &  12.2 &  33.8 &  11.7 & 50 &   Fe XII & 312.25 & 6.3 &  ---  &  ---  &   3.6 &  ---  \\
     13 &  Fe XIII & 288.57 & 6.3 &  ---  &  ---  &   5.5 &  ---  & 51 &  Co XVII & 312.54 & 6.8 &  15.7 &   5.6 &  13.5 &   8.5 \\
     14 &   Fe XIV & 289.15 & 6.3 &   8.2 &   4.4 &  14.7 &   8.1 & 52 &    Fe XV & 312.56 & 6.4 &  ---  &   4.2 &  20.5 &   6.5 \\
     15 &  Ni XIII & 290.57 & 6.3 &   6.5 &  ---  &   7.3 &  ---  & 53 &  Fe XIII & 312.87 & 6.3 &  ---  &  ---  &   8.2 &   6.8 \\
     16 &   Fe XIV & 290.74 & 6.3 &   4.2 &   5.0 &   8.2 &  ---  & 54 &  Mg VIII & 313.74 & 6.0 &  12.6 &  11.2 &  17.7 &   8.9 \\
     17 &   Fe XII & 291.01 & 6.3 &  14.8 &  10.6 &  16.5 &  10.8 & 55 &  Si VIII & 314.36 & 6.1 &   6.1 &  ---  &   8.0 &   4.7 \\
     18 &   Fe XIV & 291.60 & 6.3 &  14.7 &   9.0 &  24.2 &  12.6 & 56 &  Mg VIII & 315.02 & 5.9 &  33.7 &  22.9 &  48.7 &  26.7 \\
     19 &     S XI & 291.81 & 6.3 &  16.7 &  ---  &  ---  &   3.2 & 57 &  Si VIII & 316.22 & 5.9 &   8.8 &   6.3 &  20.9 &   6.4 \\
     20 & Ni XVIII & 291.98 & 6.6 & 138.0 &  73.7 & 245.0 &  85.3 & 58 &   Ni XIV & 316.25 & 6.3 &   4.2 &  ---  &   5.3 &  ---  \\
     21 &   Fe XIV & 292.07 & 6.3 &  ---  &  ---  &   9.0 &  ---  & 59 &   Fe XII & 316.85 & 6.3 &   3.8 &  ---  &   4.5 &  ---  \\
     22 &    Fe XV & 292.27 & 6.3 &  10.7 &   5.8 &   9.6 &   3.9 & 60 &  Mg VIII & 317.03 & 5.9 &   4.7 &   4.3 &   9.6 &   5.8 \\
     23 &  Fe XXII & 292.46 & 7.1 &  14.7 &  11.7 &  18.2 &   4.0 & 61 &  Fe XIII & 318.13 & 6.3 &  14.0 &   7.6 &  17.5 &   8.7 \\
     24 &    Si IX & 292.76 & 6.1 &   5.8 &   4.2 &   8.5 &   4.0 & 62 &   Mg VII & 319.03 & 5.8 &   5.5 &   5.4 &  13.7 &   7.0 \\
     25 &    Si IX & 292.81 & 6.2 &   5.9 &   4.3 &   8.7 &   4.1 & 63 &  Si VIII & 319.84 & 6.1 &  14.3 &  12.0 &  29.7 &  15.9 \\
     26 &    Si IX & 292.86 & 6.1 &   5.1 &   3.7 &   7.5 &   3.5 & 64 & Ni XVIII & 320.57 & 6.8 &  78.8 &  34.5 &  98.2 &  52.8 \\
     27 &   Ni XII & 295.32 & 6.3 &   6.2 &  ---  &   5.8 &  ---  & 65 &  Fe XIII & 320.81 & 6.3 &  19.3 &  12.3 &  32.8 &  13.8 \\
     28 & Fe XVIII & 295.68 & 6.9 &  10.0 &   4.6 &   6.7 &   6.0 & 66 &  Fe XIII & 321.40 & 6.3 &   3.1 &  ---  &   4.4 &   6.4 \\
     29 &    Si IX & 296.11 & 6.2 &  23.6 &  14.6 &  37.3 &  14.5 & 67 &    Fe XV & 321.77 & 6.3 &   5.2 &   3.8 &  10.6 &   5.8 \\
     30 &    Si IX & 296.21 & 6.2 &   6.8 &   4.2 &  10.7 &   4.2 & 68 &   Fe XII & 323.41 & 6.3 &   6.8 &   6.9 &  11.6 &   8.7 \\
     31 &    Ni XV & 298.15 & 6.4 &   5.9 &   3.5 &  12.8 &   4.7 & 69 &  Fe XVII & 323.65 & 6.9 &   4.9 &   4.0 &  14.3 &   7.0 \\
     32 &   Fe XIV & 299.41 & 6.3 &   8.4 &   4.0 &   5.2 &   3.7 & 70 &    Fe XV & 324.98 & 6.3 &  ---  &  ---  &   4.5 &  ---  \\
     33 &    S XII & 299.54 & 6.3 &  10.8 &   9.6 &  16.7 &   4.9 & 71 &   K XVII & 326.78 & 7.0 &   4.3 &   5.2 &   8.4 &   5.2 \\
     34 &    S XII & 299.78 & 6.4 &  ---  &   3.0 &  ---  &  ---  & 72 &    Fe XV & 327.03 & 6.4 &   9.5 &   7.0 &  15.4 &   7.2 \\
     35 &   Cr XIV & 300.30 & 6.3 &   3.0 &  ---  &   5.1 &  ---  & 73 &  Cr XIII & 328.27 & 6.3 &   9.7 &  12.8 &  16.7 &  16.0 \\
     36 & Ca XVIII & 302.19 & 7.0 &  99.9 &  63.6 & 188.0 &  59.7 & 74 &    Fe IX & 329.90 & 5.9 &  ---  &  ---  &   3.4 &  ---  \\
     37 &   Ni XIV & 302.27 & 6.3 &   6.2 &  ---  &   4.6 &  ---  & 75 &     Al X & 332.79 & 6.2 &  11.4 &   4.9 &  ---  &  11.9 \\
     38 &    Fe XV & 302.33 & 6.3 &  12.1 &   5.7 &   9.7 &   3.7 &    &          &        &     &       &       &       &       \\ 

  \enddata
  \tablecomments{Columns correspond to different
  flares, rows denote different spectral lines. The flux in a particular spectral
  line corresponds to the whole flaring region (and do not contain $\frac{1}{sr}$ factor).
	  Minus '--' sign denotes that the line was too weak in a particular spectrum.}
  \end{deluxetable}

In the {\bf V190} channel the strongest lines are: \ion{Fe}{10} 177.25~\AA,
\ion{Fe}{11} 180.41~\AA, selfblend \ion{Fe}{12} 186.85+.89~\AA, selfblend
\ion{Fe}{11} 188.23+.29~\AA, \ion{Fe}{24} 192.03~\AA, \ion{Fe}{12} 192.39~\AA,
blend of \ion{Fe}{11} 192.83 + \ion{Ca}{17} 192.85~\AA, \ion{Fe}{12}
193.51~\AA, \ion{Fe}{12} 195.12~\AA, \ion{Fe}{13} 196.54~\AA, \ion{Fe}{12}
196.63~\AA, \ion{Fe}{13} 200.02~\AA, \ion{Fe}{13} 202.04~\AA, selfblend
\ion{Fe}{13} 203.80+.83~\AA, which have intensities of order $2 \cdot 10^{-4}
\mbox{ erg/s/cm$^2$}$ and higher.  

In the {\bf U304} channel the strongest lines
are: \ion{Fe}{15} 284.16~\AA, blend \ion{S}{12} 288.42~\AA + \ion{Fe}{14}
289.15~\AA, \ion{Ni}{18} 291.98~\AA, selfblend \ion{Si}{9} 296.11+.21~\AA,
\ion{Ca}{18} 302.19~\AA, blend \ion{Fe}{17} 304.82~\AA + \ion{Fe}{15}
304.89~\AA, \ion{Mg}{8} 315.02~\AA, blend \ion{Ni}{18} 320.57~\AA +
\ion{Fe}{13} 320.81~\AA, which have intensities of order $2 \cdot 10^{-4}
\mbox{ erg/s/cm$^2$}$ and higher. The strongest line in the spectral region ---
the \ion{Si}{11}/\ion{He}{2} blend with $\lambda \sim 304$~\AA\ was removed
from the spectra before the analysis.

Emission of hot spectral lines such as \ion{Fe}{24} 192.03~\AA\
($T_{m}=16$~MK), \ion{Ca}{17} 192.85~\AA\ ($T_{m}=6.3$~MK), \ion{Fe}{22}
292.46~\AA\ ($T_{m}=13$~MK), \ion{Ca}{18} 302.19~\AA\ ($T_m=10$~MK),
\ion{Fe}{20} 309.29~\AA\ ($T_m=10$~MK) is produced only during flares. Spectral
images of a flare in these lines are compact and usually not intermingled with
other spectral lines (we inspected a large number of the SPIRIT
spectroheliograms). Thus, these lines can be used for detection of a solar
flare and they are ideal for high-temperature DEM and Doppler shift analysis.

The obtained spectra of all flares are similar, but still there are some
differences. Absolute fluxes in separate spectral lines measured by SPIRIT in
the M5.6 and X3.4 are similar and twofold higher than those in the X1.3 and X17
flares. The decrease is in direct correlation with the decrease of the total flux in
the EIT images. The decrease may be caused by variation in solar irradience ---
the M5.6 and X3.4 flares were registered at the end of 2001 (near the maximum of
solar activity), the X1.3 was observed on July 2004, and the X17 flare was
observed on September 2005 (near the minimum of solar activity), as well as
degradation of EIT sensitivity \citep{BenMoussa2013}.

\subsection{Plasma diagnostics}
\begin{figure} 
  \plotone{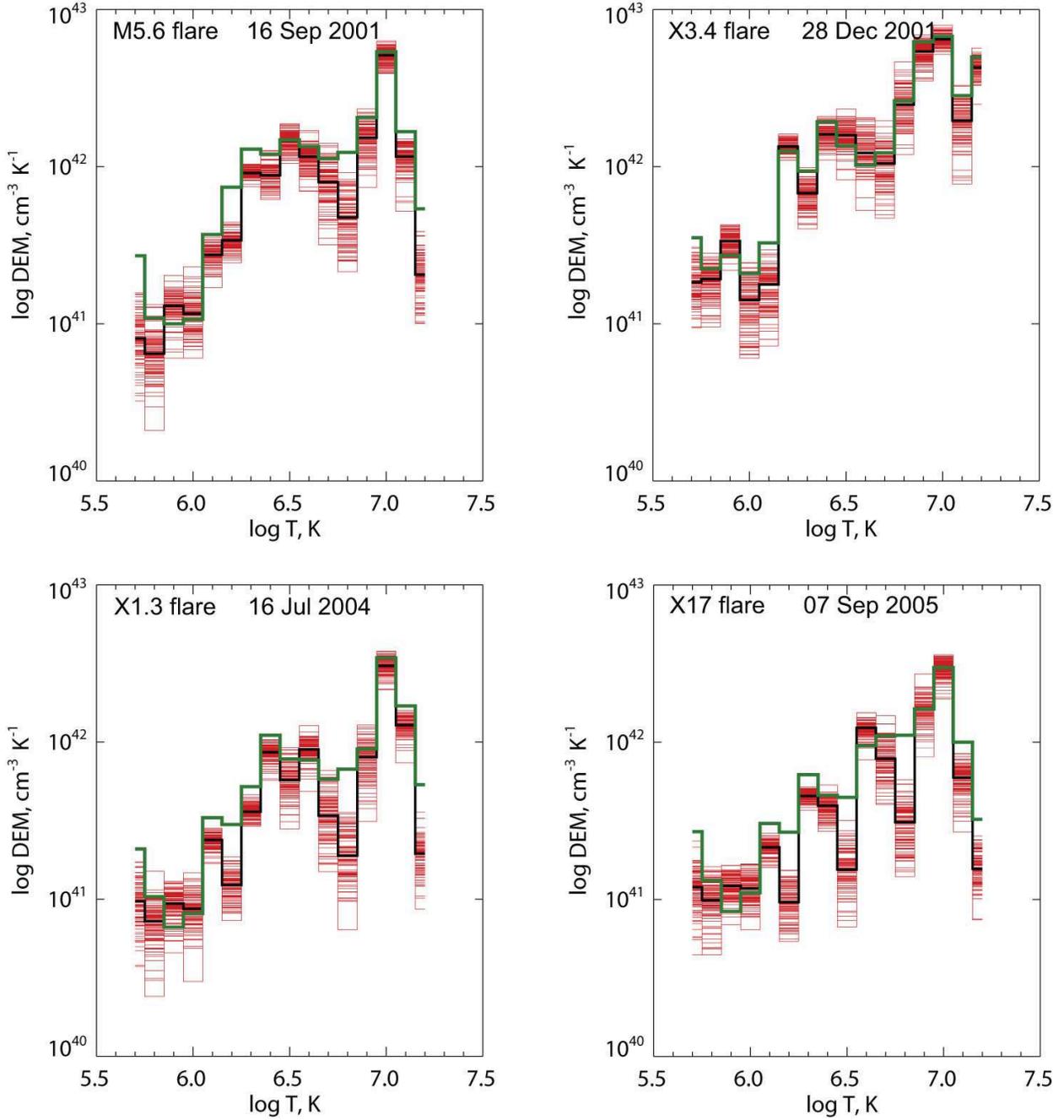}
  \caption{DEMs of the flares. Red lines denote DEMs obtained during different
  iterations of the GA procedure, black {lines} denote average value, and a dashed
  line denotes DEM, obtained during 0-level analysis.}
  \label{DEMs}
\end{figure}

The result of the DEM reconstruction is presented in Figure~\ref{DEMs}: red lines
correspond to different runs (we used 100 runs), black line is an average
(median) DEM, and a green line denotes initial DEM, obtained on the 0-level
step. 

The obtained DEMs have a similar shape --- a local minimum at $T \sim 0.6 -
0.8$~MK (cold plasma), a local maximum at $T \sim 2.5$~MK (warm plasma), and a
global maximum at $T \sim 10$~MK (hot plasma). The two-peaks shape may be
associated with different structures: the warm plasma fill loops, which are
adjacent to the flaring region \citep{Schmelz2011}, whereas the hot plasma is
produced in the flaring region. The M5.6 and X1.3 flares have narrower
hot-component peaks, which may be attributed to the earlier phases of
the flare decays ($\Delta \tau \sim 10$ and $7$ minutes after the flare
maxima).  The X3.4 and X17 were registered on later phases ($\Delta \tau \sim
40$ and $\sim 3$ hours after the flare maxima), therefore the hot plasma had
time to warm up the surroundings. The warm-component peak in
the latter two flares has two-peaks shape with $T_1=1.6$~MK (both flares) and
$T_2=2.5$~MK (the X3.4 flare) and $T_2=4.0$~MK (the X17 flare). These double
peaks in warm plasma may also be attributed to spatially separated structures.

The steep decrease in DEMs with $T>10$~MK observed in the M5.6, X1.3 and X17 flares
is determined by the intensities of hot lines, among which are \ion{Ca}{17}
($T_{m}=6.3$~MK), \ion{Fe}{22} ($T_{m}=13$~MK), \ion{Ca}{18} ($T_m=10$~MK),
\ion{Fe}{20} ($T_m=10$~MK),  but the primary contribution is definitely  due to
the \ion{Fe}{24} 192.03~\AA\ line, which has $T_{m}=16$~MK. Since the {\bf
V190} channel observations were unavailable for the X3.4 flare, it is possible
that  DEM values with $T>10$~MK are overestimated in the flare. 

The confidence level of the DEMs is assessed by the relative spread of
different DEM solutions and amounts as much as a factor of 2 (each solution from the
range equally well describes the observational data, so each solution from the
range is equally possible).

%\placefigure{DEMs}

\begin{figure}
    \includegraphics[scale=0.8]{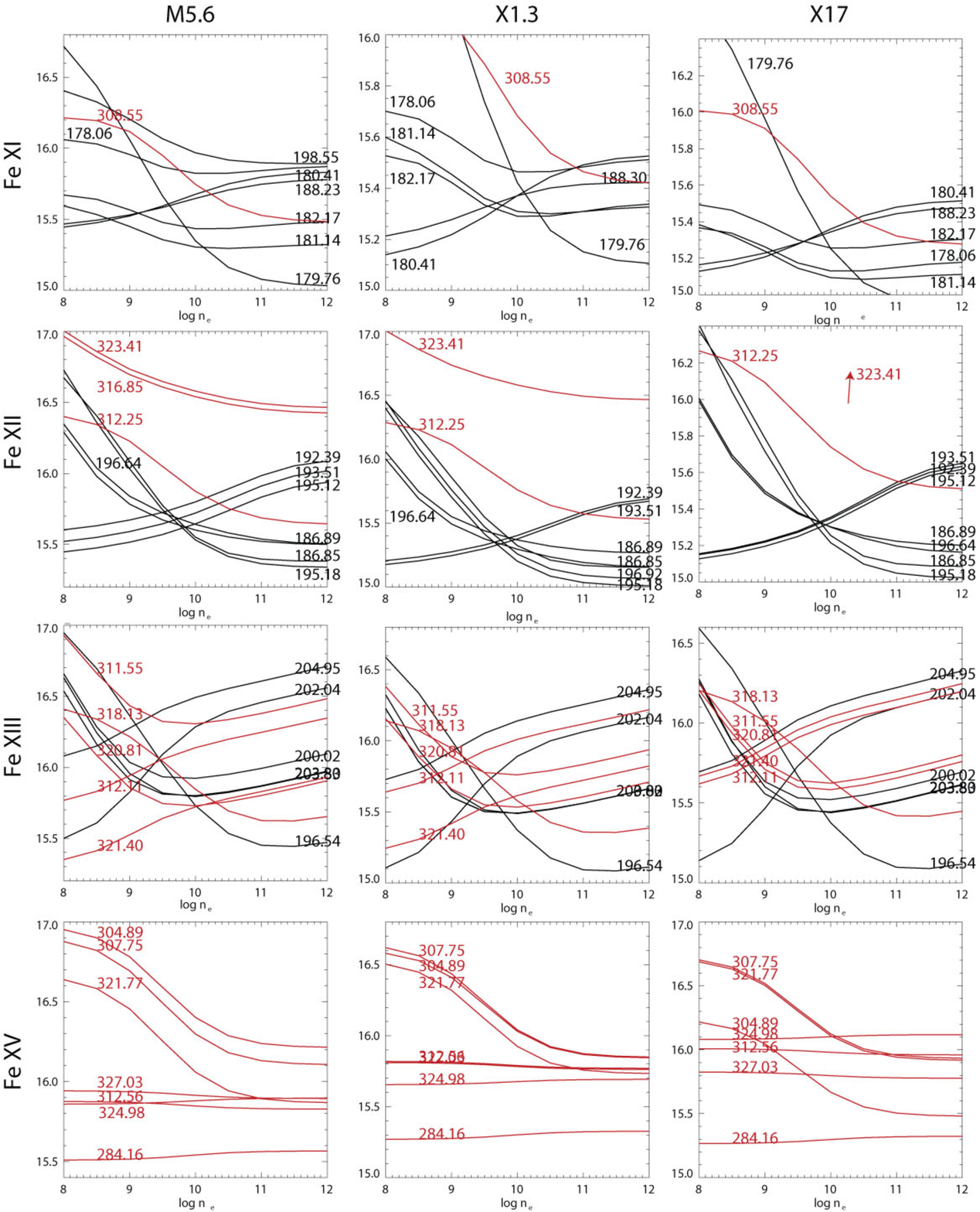}
    \caption{The L-function of \ion{Fe}{11}--\ion{Fe}{13}, and \ion{Fe}{15} ions
    for the M5.6, X1.3 and X17 flares. Black lines denote spectral lines from the {\bf V190}
      channel; red lines denote the {\bf U304} channel.}
      \label{ne_common}
  \end{figure}

The results of $n_e$ analysis are presented on Figure~\ref{ne_common} --- the
L-functions of the \ion{Fe}{11}, \ion{Fe}{12}, \ion{Fe}{13} and \ion{Fe}{15}
ions for the M5.6, X1.3 and X17 flares are given (black lines denote {\bf V190}, red lines
denote {\bf U304}). Whereas L-functions of a single ion should cross in a single
point, one can see considerable disagreement in several cases. We note that
during the iterative steps we tried to improve the agreement of the L-functions
varying the intensities of blended lines; however, better agreement was not
reached.

The best consistency among different spectral lines is observed in the
\ion{Fe}{12} ion. The $n_e$ values are $\sim 6.5 \times 10^9$~\cmt\ (the
L-functions intersect in the range $\log n_e = [9.7-9.9]$) in all flares. The
most reliable \ion{Fe}{11} lines --- 179.76, 180.41, 188.23~\AA\ --- also favour 
this value. The L-functions of the \ion{Fe}{13} ion show considerable
discrepancy. The 200.02, 202.04, and the blend $203.80+.83$~\AA\ lines show
systematically lower density $n_e \sim 2 \times 10^9$~\cmt, whereas 196.54 and
200.02, $203.80+.83$~\AA\ lines cross at density $\sim 6.5 \times 10^9$~\cmt. The
\ion{Fe}{15} lines show systematic discrepancies in all flares. We will discuss
possible causes of the discrepancies in the next section. We used a value of $n_e=6.5 \times
10^9$~\cmt\ for all the flares for DEM analysis and calculation of synthetic
spectra.

  %\placefigure{ne_010916}

\subsection{Comparison of observational and theoretical line intensities}
We compared the observational and theoretical intensities using different
approaches: in the DEM reconstruction procedure, by using L-function plots, and by
comparing observational and synthetic spectra. All these approaches, in essence,
consist of comparison of observational and theoretical line intensities, whereas
each approach gives some additional convenience in data analysis. 

In the vast majority of spectral lines the correspondence of observational and
theoretical intensity is within factor of 2. The most striking discrepancy $R$
(ratio of observational/theoretical intensity) is observed in the DEM
reconstruction in the following lines: \ion{Fe}{12} 323.41~\AA\ ($\sim 8$),
\ion{Fe}{13} 312.11~\AA\ ($\sim 2.5$), \ion{Fe}{12} 312.25~\AA\ ($\sim 2.5$),
\ion{Fe}{13} 202.04~\AA\ ($\sim 2.0$), \ion{Fe}{17} 323.65~\AA\ ($\sim 0.7$),
\ion{Fe}{15} 327.05~\AA\ ($\sim 0.7$). There is also a systematic discrepancy in
relative intensities of \ion{Mg}{8} and \ion{Si}{8} lines --- whereas these
spectral lines have similar dependence on temperature and density, the ratio $R$
for \ion{Mg}{8} is constantly higher and for \ion{Si}{8} is constantly lower
than 1. The L-function plots show discrepancies in the \ion{Fe}{13} and
\ion{Fe}{15} lines. Comparison of the observational and synthetic spectra reveal
several discrepancies in other lines. The observed discrepancies are typical for
the analyzed flares, and we will discuss them all together. 

The observed intensity of the \ion{Fe}{12} 323.41~\AA\ line is approximately
8 times higher than those, predicted in the DEM reconstruction. The discrepancy
can not be attributed to problems with SPIRIT spectral sensitivity due to 
good correspondence of other intense lines with close wavelengths. The
\ion{Fe}{12} 323.41~\AA\ line is blended with \ion{Fe}{17} 323.65~\AA; however,
the spectral profile of the blend seems unlikely to the blend of two close
spectral lines. The incorrect identification of one of the lines seems quite
reasonable.  

The next two lines with ratio $R \sim 2.5$ are \ion{Fe}{13} 312.11~\AA\ and
\ion{Fe}{12} 312.25~\AA. The lines  fall within a wide blend, which
encompassess \ion{Fe}{13} 311.55~\AA, \ion{Ni}{15} 311.76~\AA,
\ion{Mg}{8} 311.76~\AA, \ion{Fe}{13} and \ion{Fe}{12}, \ion{Co}{17} 312.54~\AA,
\ion{Fe}{15} 312.56~\AA, and \ion{Fe}{13} 312.87~\AA\ (the strongest lines
according to the synthetic spectra). Detailed analysis of the blend
deserves effort and attention; a quick look (using both spectra in
Figure~\ref{U304_flares} and L-functions in Figure~\ref{ne_common}) shows
no simple solution for improving the $R$ ratios neither via changes of
$n_e$, nor changes of relative intensity of lines involved in the blend.
Fortunately, in the {\bf U304} channel there are a number of strong lines
suitable for reconstruction of intensities of \ion{Mg}{8} and \ion{Fe}{13}
spectral lines. 

The \ion{Fe}{13} 202.04~\AA\ line is among the most intense lines of the
\ion{Fe}{13} ion; however, its ratio $R$ is approximately 2 in all analyzed
flares.  The observed discrepancy {cannot} be attributed to problems with
SPIRIT spectral sensitivity, since the line falls between other strong
\ion{Fe}{13} lines --- 200.02~\AA\ and a blend of $203.80+.83$~\AA\ (observed
intensity of these lines is consistent with theory). During the DEM
reconstruction other \ion{Fe}{13} lines (196.54, 200.02, 203.80, 312.11, 318.13,
and 320.81~\AA) were predicted with higher accuracy (usually better
than 40\%), eliminating possible issues with abundances or temperature
distribution of the emitting plasma. The L-functions of all \ion{Fe}{13} lines
are density-sensitive (see Figure~\ref{ne_common}) and correction of
the $n_e$ value seems reasonable and sufficient. However, the L-functions of
the 200.02, 203.80, 203.83, and 320.81~\AA\ lines have the same dependence on
density and their absolute values are in a good agreement with theory. Density
values $n_e$ obtained by the crossing of the 202.04~\AA\ line and 200.02 and
$203.80+.83$~\AA\ lines, is systematically lower ($\sim 2 \times 10^9$~\cmt)
than $n_e$ obtained with the \ion{Fe}{11} and \ion{Fe}{12} ions, which favors
\emph{against} the 202.04~\AA\ line. A good agreement with the density $n_e$,
obtained with the \ion{Fe}{11} and \ion{Fe}{12} ions, was obtained by
crossing the L-functions of the 196.54~\AA\ line with the 200.02,
$203.80+.83$~\AA\ lines. This result is in a slight contradiction with
\citet{Brosius98} and \citet{Shestov2009e}, who found good correspondence of the
$n_e$ values measured by \ion{Fe}{11} and \ion{Fe}{13} lines. The L-functions of
the 204.95, 312.11, and 321.40~\AA\ lines have similar behavior with density,
and in some flares their absolute values are in a good agreement.  However, the
L-function of the 204.95~\AA\ line does not produce a reasonable $n_e$
value (observed $n_e < 10^9$~\cmt). In two flares (M5.6 and X1.3) the L-function
of the 321.40~\AA\ line has a common crossing with the 196.54, 200.02, and
$203.80+.83$~\AA\ lines, which confirms the correctness of the latter lines.

Given the above information, overall agreement of the \ion{Fe}{13} L-functions
may be improved by decreasing the L-functions of the 202.04~\AA\ line by factor
of $\sim 2$ and of the 204.95~\AA\ line by factor of $\sim 4$. The observed
excess in the L-functions may be caused by unaccounted blending of spectral
lines or inappropriate atomic data. The atomic structure of the \ion{Fe}{13} ion
has recently been extensively studied by \citet{DelZanna2011b}, and the author
did not find any problems with the ion. However, observed discrepancies in the
SPIRIT data (with no strong blend candidates provided by CHIANTI) and
inconsistency in the $n_e$ values obtained with \ion{Fe}{13} indicate that some
questions still remain.  

The \ion{Fe}{17} 323.65~\AA\ line has a ratio $R \sim 0.7$ and is blended with
\ion{Fe}{12} 323.41~\AA. The latter line shows a striking discrepancy with
$R \sim 8$). The spectral profile of the blend seems unlikely to the
blend of two close spectral lines, and wrong identification of one of the lines
seems quite reasonable.   

The \ion{Fe}{15} 327.03~\AA\ line has a ratio $R \sim 0.7$. It is blended with
\ion{K}{17} 326.78~\AA\, and taking into account that both lines are on the edge
of the SPIRIT spectral range, the ratio $R$ is not too bad.

The other issue is the systematic discrepancy of intensities between the
\ion{Mg}{8} and \ion{Si}{8} spectral lines. The two ions have similar atomic
structure ($2s 2p^2 \rightarrow 2s^2 2p$ transitions in  \ion{Mg}{8} and $2s
2p^4 \rightarrow 2s^2 2p^3$ transitions in \ion{Si}{8}), similar abundances
(both in coronal and photospheric models), close wavelengths --- 311.77, 313.74, 315.02,
and 317.03~\AA\ (\ion{Mg}{8}) and 314.36, 316.22, 319.84~\AA\ (\ion{Si}{8}) ---
and contribution functions of the lines have similar dependence on temperature and
density. Nevertheless, ratios $R$ for \ion{Mg}{8} lines are
1.6, 1.4, and 1.03 for the 313.74, 315.02, and 317.03~\AA\
lines (averaged by flares), whereas the ratios $R$ for \ion{Si}{8} approach 0.7, 0.6, and 0.8 for the
314.36, 316.22, 319.84~\AA\ lines. Inadequate abundances are the most likely
cause of the discrepancy. A similar possibility was pointed out by
\citet{Schmelz2012} {in their analysis of} SERTS data.

The L-functions of the \ion{Fe}{15} ion can be separated into two groups: those
that decrease with density (the 304.89, 307.75, and 321.77~\AA\ lines) and those
that do not change with density (the 284.16, 312.56, 324.98, and 327.03~\AA\
lines).  The L-functions inside each group should coincide.  According to the
observational data, the L-function of the 284.16~\AA\ line is usually 4 times
lower {than} the others. The discrepancy can be caused by different factors: the
SPIRIT detector saturation, problems with SPIRIT spectral sensitivity, optical
thickness of the emitting plasma, and others. That is why the 284.16~\AA\ line
was not taken into account in the DEM reconstruction. The two L-function groups
are likely to cross at densities $n_e> 10^{10}$~\cmt\ (higher {than} the density
obtained by the \ion{Fe}{11} and \ion{Fe}{12} ions).

\begin{figure} 
  \plotone{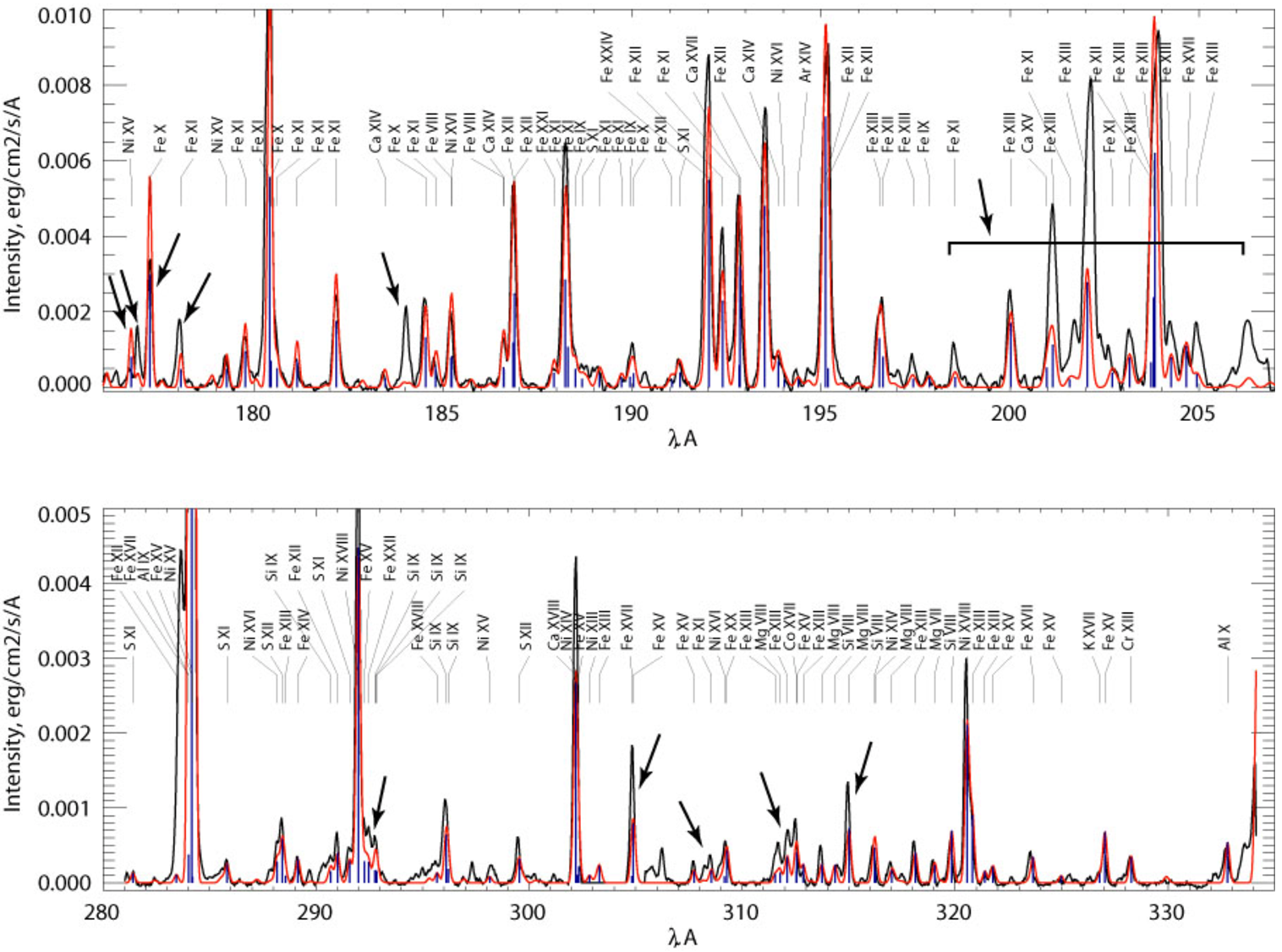}
  \caption{Comparison of observational and synthetic (calculated using DEM and
  $n_e$ with CHIANTI) spectra for the M5.6 flare. Top panel ---  {\bf V190}
  channel, bottom panel --- {\bf U304} channel. The black curve denotes
  observational data, blue vertical lines denote individual spectral lines, and
  {the} red curve denotes synthetic spectra (calculated using instrument FWHM). The
  arrows show major discrepancies of observational and synthetic spectra.}
  \label{Exp_vs_synth}
\end{figure}

The other method for analyzing the correspondence of observational data with
theoretical values is comparison with unmodified synthetic spectra --- that was
calculated using DEM and $n_e$ (and other model parameters) and \emph{has not}
been modified to match observational data. Such a comparison for the M5.6 flare
is given in Figure~\ref{Exp_vs_synth}. The main discrepancies are observed in
the {\bf V190} channel, with the $R$ ratios of different lines being both larger
and smaller than the unity: \ion{Ni}{15}~176.74~\AA\ ($R \ll 1$),
\ion{Ca}{15}~176.93~\AA\ ($R \gg 1$), \ion{Fe}{10}~177.24~\AA\ ($R \sim 0.5$),
\ion{Fe}{11}~178.06~\AA\ ($R \sim 2$), line with $\lambda \sim 186.00$~\AA\
(candidates are \ion{Ni}{15}~185.73~\AA, \ion{Fe}{13}~185.76~\AA\ and
\ion{Fe}{12}~186.24~\AA, which all have $R \ll 1$). Many spectral lines of the
\ion{Fe}{13} ion with wavelengths $\lambda \sim 200-205$ deviate from
theoretical values. In the {\bf U304} channel the spectral lines are more or
less consistent with the theory, with the {following} exceptions:
\ion{Fe}{15} 284.16~\AA\ ($R \sim 0.25$), \ion{Fe}{15} 292.28~\AA\ ($R \sim
0.5$), blend of \ion{Fe}{17} 304.82 + \ion{Fe}{15} 304.89~\AA\ ($R \sim 0.5$),
\ion{Si}{8} 308.19~\AA\ ($R \sim 3$), complex blend with $\lambda \sim 312$~\AA\
(\ion{Mg}{8}, \ion{Fe}{13}, \ion{Fe}{12}, \ion{Co}{17} and others ions),
\ion{Mg}{8}~315.02~\AA, \ion{Fe}{12}~323.41~\AA.

%\placefigure{Exp_vs_synth}

\section{Discussion}
During the identification we fit observational data with synthetic
{spectra} calculated using CHIANTI. The procedure does not take into
account unknown lines, and {this} is the main disadvantage of the
proposed method of the identification.  However, comparison of observed and
calculated intensities {gives} a lot of information about the reliability
(qualitative --- correct or incorrect) and accuracy (quantitative --- say 10\%
or 50\%) of the identification. The accuracy depends on two factors --- a)
{quality} of the observational data (accuracy of spectral calibration,
low noise, absence of scattered light in the instrument, the compactness of the
emitting plasma etc.); b) accuracy of the used atomic data. 

We analyzed how the identification of the obtained spectra depended on the
relative calibration of SPIRIT and conclude that the relative calibration is
better than a factor of 2 (there are a number of lines whose intensity ratio do not
depend on density and which comply with theory). The absolute calibration was obtained
using simultaneous EIT images --- 195~\AA\ for the {\bf V190} channel and
304~\AA\ for the {\bf U304} channel. After the absolute calibration, the average
line intensities in the both channels (which were calibrated independently)
satisfy each other. However, the resulting absolute calibration of
SPIRIT is as good as EIT calibration. Any errors in EIT calibration --- for example due
to the decay of the EIT sensitivity \citep{BenMoussa2013} --- will affect the
absolute calibration of the presented spectra.

We analyzed how the results of the DEM reconstruction depend on
the errors (up to 30\%), artificially introduced {into} the observational data.
The obtained discrepancy turned out to be within the DEM confidence level, obtained
in the genetic algorithm. Other factors (beside adequate identification) do not play such an
important role {in} the final accuracy. We assess the final accuracy of the
observational data (including the absolute calibration) to be a factor of 4.

The performed analysis of simultaneous EIT and SPIRT data proved that telescopic
and spectroscopic observations significantly enhance each other. SPIRIT gives
direct information about the relative flux in each spectral line contributing to
the EIT image, and relative flux measured by EIT (in dn units) allows to
{calibration of} uncalibrated SPIRIT data. {Using spectroscopic
instrumentation} with relatively high spectral resolution could enhance
{the} informational content of other instruments, like AIA or EVE.

The other important aspect for spectroscopic analysis is spectral resolution.
The instrumental resolution of SPIRIT is $\sigma \sim 0.1$~\AA\ (for comparison,
the spectral resolution of EIS is $\sigma \sim 0.020$~\AA). Many
strong and important lines are not resolved by SPIRIT, such as the \ion{Fe}{13}
203.80 and 203.83~\AA\ lines, or \ion{Ni}{18}~291.98~\AA,
\ion{Fe}{15}~292.26~\AA, and \ion{Fe}{22}~292.45~\AA\ lines. Nevertheless, the
identification procedure used, allowed the deconvolution of blends and the
calculation of intensities. In some cases we obtained good correspondence of
observational and theoretical intensities: the example is the blend of the
\ion{Fe}{13}~203.80 and 203.83~\AA\ lines, which are in a good agreement with
the \ion{Fe}{13}~200.02~\AA\ line; the blend of the \ion{Fe}{13}~196.54~\AA\ and
\ion{Fe}{12}~196.64~\AA\ lines also complies {with} the rest of the \ion{Fe}{12} and
\ion{Fe}{13}. In some cases agreement was not achieved: the multiple blend with
$\lambda \sim 312$~\AA\ is an example, where the lines contributing to
{the} blend show poor correspondence with the theory. Nevertheless, the method of
deconvolving blends using a synthetic spectrum is a powerful tool for
spectroscopic analysis.

Let's compare the calculated DEMs with those obtained from other instruments.
The DEM which is widely used for modelling of the EUV spectra is that presented
by~\citet{DereCook79} (this DEM is actually provided by CHIANTI as a
\texttt{flare.dem}). The authors analyzed {the} decay phase of {an}
M2 flare using observations from the S082A EUV spectroheliograph and the S082B
UV spectrograph aboard Skylab. During the DEM calculation the authors used the
quantity ``total line power radiated by the plasma'' (which actually 
coincides with our approach). However, the DEM values provided by CHIANTI are
expressed in units cm$^{-5}$ K$^{-1}$. The DEMs, calculated in our analysis
correspond to the whole flaring region, and we need to assess the area
associated with the flare. In order to assess the area we inspected
monochromatic images of the flares and conclude that the images have symmetrical
"Gaussian" shape with typical FWHM $\sim 8$~pixels (good examples are the
bright lines of the \ion{Ni}{18}, \ion{S}{12}, \ion{Ca}{18} ions in
Figure~\ref{U304_overview}).  This is the minimal size of the structure observed
on the spectroheliograms, and the size is determined by the PSF of the
instrument (primarily, its focusing mirror). Thus, the spatial size of a flare
should not exceed $\sim 3\times 3$ pixels so as not to increase flare
images. Taking into account the angular size of a pixel of SPIRIT
--- 6.7~arcsec, we obtain that $3\times 3$ pixels correspond to the area $2
\cdot 10^{18}$~cm$^2$. We multiply  \texttt{flare.dem} from CHIANTI by this
factor and compare it with the current DEMs. The correspondence is good enough:
beside similar shape (local minima and maxima coincide in the two datasets) we
get compliance of the orders of magnitude --- $\sim 8 \times 10^{40}$~\cmt\
K$^{-1}$ (cold component) and $\sim 6 \times 10^{42}$~\cmt\ K$^{-1}$ (hot
component).

We performed another verification of the calculated DEMs: we simulated GOES
X-ray fluxes using the calculated DEMs and GOES response functions
(\texttt{goes\_resp2.dat} from Solar Software). The calculated fluxes complied
within an order of magnitude or better with those, actually measured by GOES.

To investigate heating dynamics in observed flares --- impulsive or
continuous ---  we {compared} hot component lifetime ($\tau_{life}$, few hours)
with its conductive cooling time ($\tau_{cond}$): if $\tau_{cond} \ll
\tau_{life}$, then heating is continuous; if $\tau_{cond} \sim \tau_{life}$, then
heating is impulsive. We {estimated} $\tau_{cond}$ {with the formula} \citep{Culhane94}:
\begin{equation}
	\tau_{cond}=\frac{21 n_e k_b h^2}{5 \kappa T^{5/2}}
\end{equation}
where, $n_e$ --- plasma electron density, $k_b$ --- Boltzmann constant, $h=1.5
\times 10^7$~m --- characteristic size, $\kappa = 9.2 \times
10^{-7}$~erg~s$^{-1}$~cm$^{-1}$~K$^{-7/2}$ --- the Spitzer conductivity,
$T=10$~MK --- plasma temperature.  Electron density in flares ranges from
$10^{10}$~\cmt  to $10^{12}$~\cmt \  \citep{Milligan2012ne}. For $n_e =
10^{10}$~\cmt \ $\tau_{cond} \sim$~30~seconds ($\tau_{cond} \ll \tau_{life}$),
which requires continuous heating; for $n_e = 10^{12}$~\cmt \ $\tau_{cond} \sim$~1~hour
($\tau_{cond} \sim \tau_{life}$), which favors impulsive heating. SPIRIT spectra
don't have high-temperature spectral lines, suitable for the density diagnostics, so we estimate
$n_e$ using the obtained DEMs:
\begin{equation}
EM_{hot} \sim n_e^2 h^3 \Rightarrow 
n_e \sim \sqrt{\frac{EM_{hot}}{h^3}} \sim
10^{11} \ \textmd{\cmt}
\end{equation}
This is a rough estimation: the DEMs accuracy is a factor of 4, and the real volume
of the hot component is probably less than $h^3$ (most likely it is not
spherical, {but has} loop-like geometry). So $n_e$ is probably 
closer to $10^{12}$~\cmt \ than to $10^{10}$~\cmt, and heating in observed flares is
most likely impulsive.

\section{Conclusion}
Initially, the main goal of the work was to present unique observational data
--- EUV spectra of large solar flares, observed by the SPIRIT spectroheliograph.
Due to the relatively low spectral resolution of SPIRIT, many lines are
blended, which prevents a straightforward method for line identification and
measurement. The original procedure for spectra analysis, based on calculation
of synthetic spectra and measurement of plasma DEM and $n_e$, not only
allowed identification and measurement of intensity of as many as 70
spectral lines in each spectral band in each flare, but also provided a lot of
other important information. The performed spectroscopic analysis
demonstrated the accuracy of the adopted spectral calibration of the
SPIRIT spectroheliograph.  Simultaneous observations of the EIT telescope and
the SPIRIT spectroheliograph allowed calculation of \emph{absolute} fluxes in each
spectral line. 

Whereas the analysed flares belong to different X-ray classes and
were registered on different stages of their decay, registered spectra and
calculated DEMs have many in common.  All DEMs have similar shape with global
maxima at $T \sim 10$~MK and local maxima at $T \sim 2.5$~MK. 

The performed comprehensive analysis allowed interpretation of
observational data with good quality --- most of the spectral line intensities
correspond to their theoretical values with 40\% accuracy. The remaining lines
with consistency of a factor of 2 and worse require additional analysis, which may
involve, along with refinement of spectral calibration, more complicated plasma
models, verification of abundances or atomic rates etc. 

The registered spectra, as well as proposed identification and
DEMs, could be used for further spectral analysis. The obtained spectra,
synthetic spectra, DEMs, and proposed IDL software are available at
\url{http://xras.lebedev.ru/SPIRIT/} or on request from S. Shestov.

\acknowledgments
We are grateful to Sergey Bogachev and Giulio Del Zanna for discussions, comments
and invaluable help. We thank the anonymous referee, whose kind remarks and
suggestions helped us to improve our work.

The work was partially supported by the non-commercial Dynasty foundation, by a
grant from the President of the Russian Federation (MK-3875.2011.2), by a
grant from the Russian Foundation of Basic Research (grant 11-02-01079a),
and by a program No.  22 for fundamental research of the Presidium of the
Russian Academy of Sciences.  The research leading to these results has received
funding from the European Commission's Seventh Framework Programme
(FP7/2007-2013) under the grant agreement eHeroes (project No. 284461,
\url{www.eheroes.eu}).

\bibliographystyle{apj}
\bibliography{biblio_base}  

%\clearpage

%\clearpage

%\clearpage

%\clearpage

%\clearpage

%\clearpage

%\clearpage

%\clearpage

\clearpage

\clearpage

\end{document}